\documentclass[fleqn,usenatbib]{mnras}

\usepackage{newtxtext,newtxmath}

\usepackage[T1]{fontenc}

\DeclareRobustCommand{\VAN}[3]{#2}
\let\VANthebibliography\thebibliography
\def\thebibliography{\DeclareRobustCommand{\VAN}[3]{##3}\VANthebibliography}

\usepackage{graphicx}	
\usepackage{amsmath}	
\usepackage{subcaption}
\usepackage{rotating}

\title[Optical intranight variability of blazars]
{Detection of an intranight optical hard-lag with colour variability in blazar PKS 0735+178}
\author[C. McCall et al.]
{Callum McCall,$^{1}$\thanks{E-mail: c.mccall@2017.ljmu.ac.uk}
Helen E. Jermak,$^{1}$
Iain A. Steele,$^{1}$
Shiho Kobayashi$^{1}$
Johan H. Knapen,$^{2,3}$
\newauthor
Pablo M. Sánchez-Alarcón,$^{2,3}$
\\
$^{1}$Astrophysics Research Institute, Liverpool John Moores University, Liverpool Science Park IC2, 146 Brownlow Hill, UK\\
$^{2}$Instituto de Astrofísica de Canarias, c/Vía Láctea s/n, E-38205, La Laguna, Tenerife, Spain\\
$^{3}$Departamento de Astrofísica, Universidad de La Laguna, E-38206, La Laguna, Tenerife, Spain\\
}

\date{Accepted XXX. Received YYY; in original form ZZZ}

\pubyear{2023}

\begin{document}
\label{firstpage}
\pagerange{\pageref{firstpage}--\pageref{lastpage}}
\maketitle

\begin{abstract}
Blazars are a highly variable subclass of active galactic nuclei that have been observed to vary significantly during a single night. This intranight variability remains a debated phenomenon, with various mechanisms proposed to explain the behaviour including jet energy density evolution or system geometric changes. We present the results of an intranight optical monitoring campaign of four blazars: TXS 0506+056, OJ287, PKS 0735+178, and OJ248 using the Carlos Sánchez Telescope. 
We detect significant but colourless behaviour in OJ287 and both bluer- and redder-when-brighter colour trends in PKS 0735+178. Additionally, the $g$ band shows a lag of $\sim10\,\mathrm{min}$ with respect to the $r,i,z_\mathrm{s}$ bands for PKS 0735+178 on 2023 January 17. This unexpected hard-lag in PKS 0735+178 is not in accordance with the standard synchrotron shock cooling model (which would predict a soft lag) and instead suggests the variability may be a result of changes in the jet's electron energy density distribution, with energy injection from Fermi acceleration processes into a post-shocked medium.
\end{abstract}




\section{Introduction}
At the centre of most galaxies resides a supermassive black hole (SMBH) and if subject to accretion, this SMBH can be referred to as an active galactic nucleus (AGN). A blazar is an AGN where the relativistic jet emanating from the polar regions of the black hole is oriented along our line of sight within $\lesssim 15^{\circ}$ \citep{hovatta2009}. The resulting relativistic beaming produces highly variable emission across the electromagnetic spectrum \citep{urry1998}. 

Blazars can be categorised into two classes based on the features within their optical spectra. BL Lac-types have near featureless optical spectra whereas flat spectrum radio quasars (FSRQs) show emission lines with EW $\geq5$\,\AA\ \citep{stickel1991}. The spectral energy distribution of blazars has a distinct double-hump structure with a lower-energy peak at IR to X-ray frequencies and a higher-energy peak at X-ray to HE $\gamma$-rays \citep{fossati1998}. The lower-energy peak is attributed to synchrotron emission from the jet, whereas the source of the higher-energy peak is still debated \citep{ghisellini1998,prandini2022}. This second peak can be explained through leptonic models, where the emission is attributed to lower-energy seed photons being up-scattered by relativistic electrons through inverse-Compton scattering \citep{bottcher2013b}. The source of these seed photons can be from within \cite[synchrotron self-Compton;][]{maraschi1992} or from outside \cite[external Compton;][]{dermer1993} the jet. This means relationships between the lower- and higher-energy regimes are likely, as their origin is the same population of electrons. Conversely, hadronic models involving interactions between protons and/or mesons might explain the high energy variability independently of the lower-energy emission \citep{bottcher2013b}. These hadronic models are also important for the production of HE neutrinos \citep{osorio2023}.

The location of the lower-energy peak can be used to distinguish between different blazar classes and BL Lac-type subclasses. Low synchrotron peaked (LSP) sources have a synchrotron peak $<10^{14}\,\mathrm{Hz}$, corresponding to IR emission, and can be FSRQs or BL Lacs (low-frequency peaked BL Lacs; LBLs). Only BL Lacs have synchrotron peaks at higher frequencies. High synchrotron peaked (HSP or high-frequency peaked BL Lacs; HBLs) sources have synchrotron peaks $>10^{15}$ Hz in UV or X-ray; those with a synchrotron peak frequency in the optical regime between $10^{14}\leq\nu\leq10^{15}$ Hz are classed as intermediate synchrotron peaked (ISP) sources or intermediate-frequency peaked BL Lacs (IBLs) \citep{urrypadovani1995,abdo2010a}.

Blazars can show variability on a variety of time-scales, from years down to minutes, with the latter being referred to as intranight variability. The intranight optical variability (INOV) of blazars has been the focus of numerous campaigns over the past $30\,\mathrm{yr}$ \citep{miller1989,sagar2004,chand2021}, but is still not a well-understood characteristic of their emission. There have been many models proposed to explain the different types of variability such as quasi-periodic oscillations (QPOs) \citep{jorstad2022}, micro-flares \citep{bhatta2015}, and gradual flux changes \citep{pandian2022}, observed over hour-long time-scales. These models can predict time lags between different wavebands, colour evolution, and polarization degree and angle changes. Such models include geometric changes related to the jet angle and Doppler factors of emitting blobs \citep{gopal1992}, intrinsic changes relating to particle energy distributions \citep{bachev2012,bachev2015}, and extrinsic changes including microlensing from objects outside the blazar system along our line of sight \citep{paczynski1996}. 

Colour evolution in blazars generally takes one of two forms: bluer-when-brighter (BWB) or redder-when-brighter (RWB). In general, BL Lac-type sources tend to exhibit BWB trends and where this is not the case and a RWB trend is seen, the objects tend to be host-galaxy-dominated \citep{negi2022}. FSRQs can be found showing both BWB and RWB trends, but the proportion of sources that exhibit the latter is much greater than among BL Lacs \citep{zhang2015b,negi2022}. In any case, the colour changes of blazars can be used to determine the origin of the emission and refine emission mechanisms. The BWB chromatism may arise from a number of different mechanisms. One proposed mechanism is synchrotron cooling of internal shock accelerated electrons, where higher-energy electrons cool faster making bluer light appear more variable than redder light \citep{kirk1998}. Another is the one-component synchrotron model where an increase in the energy output of the blazar increases the average particle energy and thus the frequency, making the blazar appear bluer-when-brighter \citep{fiorucci2004}. When in faint states or periods of jet quiescence, redder emission and variability from the accretion disc may become visible leading to RWB behaviour; this is true for both BL Lac and FSRQ-type blazars. A potential explanation of why more FSRQs show general RWB behaviour may come from the flattening of their spectral slope at optical frequencies from the presence of the `UV bump'. The RWB behaviour, or steepening of the composite thermal and non-thermal spectrum, could be explained if the non-thermal component had a greater contribution towards the total flux during brightening \citep{gu2006}.

Similarly to colour variability, time lags (or a lack thereof) can help constrain emission processes. Time-lags between different frequencies have been observed in numerous studies \citep{chatterjee2008,gaur2012a,liodakis2018}, but inter-band lags within the optical waveband are much less commonly observed, with the first proposed detection in 2009 \citep{wu2009}. The internal shock model \citep{kirk1998} predicts time-lags across all synchrotron emission frequencies and therefore within the optical waveband. The rate of synchrotron cooling is frequency dependent, meaning variability at higher-energy precedes that at lower energies \citep{kirk1998}. 

INOV including colour evolution and inter-band time-lags is difficult to detect for a number of reasons, both physical to the system and logistically in terms of observational cadence, and can result in data with irregular or limited time resolution \citep{bottcher2010}.

In this paper, we present the results of a five-night INOV monitoring campaign executed over the period 2023 January 15-19 on four blazars: TXS 0506+056, OJ287, PKS 0735+178 and OJ248. These sources were chosen due to a combination of their historic degrees of INOV and their observability during the campaign. The sample was kept small to ensure the data were sufficient to observe hour-long variability time-scales with high sampling.

In Section \ref{sec: obs} we describe the facilities and instruments used and photometric analysis procedures including statistical processes. In Section \ref{sec: sample} we describe the individual sources and their historical behaviour. In Sections \ref{sec: analysis}, \ref{sec: colour}, and \ref{sec: lags} we present the results of the correlation analyses including variability indicators, colour behaviour, and time-lags. In Section \ref{sec: discussion} we discuss the implication of the results from the analysis. 

\section{Observations and data reduction}

\label{sec: obs}
\subsection{Carlos Sánchez Telescope - MuSCAT2}
Observations were taken with the four-colour simultaneous imager MuSCAT2 \citep{narita2019} located on the $1.52\,\mathrm{m}$ Carlos Sánchez Telescope (TCS) at the Teide Observatory, Tenerife. The TCS has a Dall-Kirkham configuration with a focal length of $f/13.8$ and the MuSCAT2 instrument is fixed at the Cassegrain focus. MuSCAT2 achieves four-colour simultaneous observations through the use of three dichroic mirrors to separate the light into four wavelength bands (\textit{g} (400--550nm), \textit{r} (550--700nm), \textit{i} (700--820nm), \textit{$z_\mathrm{s}$} (820--920nm) where the subscript `s' here denotes the `shorter' waveband range to a traditional \textit{z} band filter) to be detected by four fast-readout PIXIS CCD cameras. The derived pixel scales of $\sim0.44\,\mathrm{arcsec/pixel}$ across each band correspond to a field of view (FOV) of $7.4\times7.4\,\mathrm{arcsec}^2$ in all filters \citep{narita2019}.

A summary of the TCS observations is presented in Table \ref{sample_data}. Each source was observed on three separate nights and, in general, observations were interleaved for two sources with a typical observing sequence of 10 frames per source with a 30-second exposure time (the longest that could be executed without autoguiding; imposed by an autoguider failure). 

\subsection{Liverpool Telescope - MOPTOP}
Supplementary observations were carried out with the MOPTOP \citep{shrestha2020} polarimeter on the fully robotic, $2\,\mathrm{m}$ Liverpool Telescope (LT) located at the Observatorio del Roque de los Muchachos, La Palma. The LT has a Ritchey-Chrétien Cassegrain design with MOPTOP fitted at one of the science fold ports. 

MOPTOP boasts a dual-beam configuration utilising a continuously rotating half-wave plate and two fast readout, very low noise CMOS cameras. Together, these allow MOPTOP to achieve high sensitivity and time resolution while minimizing systematic errors. MOPTOP has a $7\times 7\,\mathrm{arcsec}^2$ FOV \citep{shrestha2020}. The data were taken in $B$ (380-–520\,nm), $V$ (490-–570\,nm) and $R$ (580-–695\,nm) filters quasi-simultaneously (observations with different filters taken in succession).

The observations with MOPTOP were taken much less frequently than those of MuSCAT2, aiming for a few observations each night wherever possible. Observations were taken on 2023 January 15-16 before poor weather conditions in La Palma closed the observatory for the remainder of the campaign.

\subsection{Data Reduction}
The data were calibrated and analysed using the \textsc{astropy} Python package and standard differential photometry techniques. MuSCAT2 frames required bias/dark subtraction and flat fielding and this was performed using calibration frames taken before/after observations each night. The data also required a WCS fit which was performed using the \textsc{Astrometry.net} API \citep{lang2010}. MOPTOP reduction is carried out using an automated data reduction pipeline running at the telescope which provides bias/dark subtracted, flat-fielded, and WCS fitted frames \citep{smith2016}.

The MuSCAT2 photometric data were calibrated by calculating the weighted average zero point in each frame using estimates from five in-frame calibration stars with known $griz_\mathrm{s}$ magnitudes from SDSS, Pan-STARRS and APASS catalogues \citep{abdurro'uf2022,flewelling2020,henden2018}. This zero point was then used to calibrate the data for the source.

MOPTOP photometric data were calibrated using an in-frame reference star with known $BVR$ magnitudes. Polarimetric data were calibrated using observations of zero-polarized standard stars to calculate instrumental polarization values ($q_0$ and $u_0$) and polarized standard stars to find the instrumental position angle values.

Before applying statistical methods and generating light curves, the data were sigma-clipped using the Python package \textsc{astropy} as a form of quality control. In this process, data are removed if they are above or below three times the standard deviation from the median value. This is an iterative process that produces updated median and standard deviation values until no further values are removed. Applying this to the flux values at each epoch, disproportionate variability between filters caused by erroneous pixels or poor sky conditions was removed. The final amount of data removed equates to less than 10 per cent (and in most cases less than 5 per cent) per source per epoch.

We note that no correction for host galaxy contribution to the source magnitudes or colours has been applied. In general blazars of subclasses LSP and ISP outshine their host galaxies by several orders of magnitude, so significant host galaxy emission is only observed when the AGN is in a very low state in combination with deep photometric imaging \citep{nilsson2012,gaur2014,olguín-iglesias2016}. None of our sources were in such a state.


\begin{table*}
\centering
\begin{tabular}{cccccccc}
\hline
Name & $\alpha$ (J2000) & $\delta$ (J2000) & Type & $z$ & Date & $H$ & $N$ \\ \hline
TXS 0506+056 & $05^\mathrm{h}09^\mathrm{m}25\fs96$ & +05\degr41\arcmin35\farcs333 & LSP & 0.337 & 2023 Jan 15 & 5.15 & 121 \\
... & ... & ... & ... & ... & 2023 Jan 16 & 1.43 & 157 \\
... & ... & ... &... & ... & 2023 Jan 18 & 0.87 & 100 \\

OJ287 & $08^\mathrm{h}54^\mathrm{m}48\fs875$ & +20\degr06\arcmin30\farcs640 & LSP & 0.306 & 2023 Jan 15 & 7.95 & 216 \\
... & ... & ... & ... & ... & 2023 Jan 18 & 2.62 & 294 \\
... & ... & ... & ... & ... & 2023 Jan 19 & 0.29 & 17 \\

PKS 0735+178 & $07^\mathrm{h}38^\mathrm{m}07\fs394$ & +17\degr42\arcmin18\farcs998 & ISP & 0.45 & 2023 Jan 15 & 3.22 & 77 \\
... & ... & ... & ... & ... & 2023 Jan 16 & 0.63 & 25 \\
... & ... & ... & ... & ... & 2023 Jan 17 & 7.24 & 313 \\

OJ248 & $08^\mathrm{h}30^\mathrm{m}52\fs086$ & +24\degr10\arcmin59\farcs820 & FSRQ & 0.939 & 2023 Jan 16 & 0.61 & 29 \\
... & ... & ... & ... & ... & 2023 Jan 17 & 7.33 & 250 \\
... & ... & ... & ... & ... & 2023 Jan 19 & 0.74 & 66 \\
\hline
\end{tabular}
\caption{List of the blazars used in this analysis including the source Right Ascension ($\alpha$), Declination ($\delta$), type, redshift ($z$), TCS observation date, hours observed ($H$), and the number of observations ($N$) in $g, r, i, z_\mathrm{s}$ bands.}
\label{sample_data}
\end{table*}

\section{Targets}
\label{sec: sample}
\subsection{TXS 0506+056}
TXS 0506+056 is situated at a redshift of 0.337 \citep{paiano2018} and until recently was classified as an LSP BL Lac (LBL) object \citep{fan2014}. It has been proposed that TXS 0506+056 may belong to a subclass of blazars named masquerading BL Lacs, or blue FSRQs \citep{ghisellini2012,padovani2019,lewis2021}, where the Doppler-boosted synchrotron radiation in the relativistic jet is bright enough to outshine the broad line region \citep{rodrigues2019,rajagopal2020}. This would make the FSRQ appear as a BL Lac object due to the apparent lack of emission lines.

In 2017 September, TXS 0506+056 was found to be in a consistent location with the IceCube neutrino event EHE 170922A \citep{kopper2017} to within a $3\sigma$ significance level \citep{icecube2018} while in a state of heightened $\gamma$-ray activity \citep{tanaka2017}. Since then, the object has been the subject of various studies over different frequencies and time-scales \citep[see][]{keivani2018,bachev2021,acciari2022}.

We observed TXS 0506+056 over three nights, totalling $7.45\,\mathrm{h}$ in the $griz_\mathrm{s}$ filters. We obtained complementary photo-polarimetric data with MOPTOP on the LT in the $BVR$ bands to assess the polarimetric state of the source during the observing campaign. We also took \textit{Fermi} $\gamma$-ray data from the Light Curve Repository (LCR) \citep{abdollahi2023}. The \textit{Fermi} data showed TXS 0506+056 to be in a $\gamma$-ray state a little higher than its median flux level over all time, at $6.63\pm6.05\times10^{-8}\,0.1\text{--}100\,\mathrm{GeV\,ph\,cm^{-2}\,s^{-1}}$ (median level: $8.30\pm0.23\times10^{-8}\,0.1\text{--}100\,\mathrm{GeV\,ph\,cm^{-2}\,s^{-1}}$). Its optical linear polarization degree varied between roughly 9 and 16 per cent and the polarization angle varied between approximately $150^{\circ}$ and $170^{\circ}$ across all $BVR$ bands.

\subsection{OJ287}
OJ287 is located at a redshift of 0.306 \citep{sitko1985} and is a well-known LSP BL Lac-type object (LBL) \citep{nilsson2018}. OJ287 is one of the best binary supermassive black hole candidates with a double-peaked outburst period of roughly $12\,\mathrm{yr}$ \citep{sillanpaa1988,sillanpaa1996}. The observed optical outbursts date back over $130\,\mathrm{yr}$, corresponding to the interaction of the secondary black hole with the primary's accretion disc. A second, longer period of approximately 60 yrs has also been reported \citep{valtonen2006} which is thought to arise from the orbital precession inducing a precession into the accretion disc \citep{katz1997,sundelius1997} and causing a subsequent wobble in the jet angle.

We observed OJ287 over three nights, totalling $10.86\,\mathrm{h}$ in the $griz_\mathrm{s}$ filters. Additional MOPTOP photo-polarimetric data showed the optical linear polarization degree varied between roughly 10 and 25 per cent and its polarization angle varied between approximately $160^{\circ}$ and $175^{\circ}$ across all $BVR$ bands. \textit{Fermi} $\gamma$-ray data from the Light Curve Repository (LCR) \citep{abdollahi2023} showed OJ287 to be in a $\gamma$-ray state a little lower than its median flux level over all time, at $3.77\pm2.93\times10^{-8}\,0.1\text{--}100\,\mathrm{GeV\,ph\,cm^{-2}\,s^{-1}}$ (median level: $5.94\pm0.31\times10^{-8}\,0.1\text{--}100\,\mathrm{GeV\,ph\,cm^{-2}\,s^{-1}}$).

\subsection{PKS 0735+178}
PKS 0735+178 is an ISP BL Lac-type object with a disputed redshift as a result of its featureless optical spectrum. A lower limit of $z\geq0.424$ was first proposed by \cite{carswell1974} after the detection of a strong absorption feature, and has since been refined using the detection of the host galaxy using deep $I$ band imaging when the central engine was in a faint state ($z=0.45\pm0.06$ \citep{nilsson2012}), and the surrounding galaxies ($z\sim0.65$ \citep{stickel1993,falomo2021}). \cite{sahakyan2023} compare the multiwavelength data to that of TXS 0506+056, another blazar neutrino candidate and utilise conditions set out in \cite{padovani2019} to conclude that like TXS 0506+056, PKS 0735+178 may also be a masquerading BL Lac as a result of a radio power $P_{1.4\,\mathrm{GHz}} > 10^{26}\,\mathrm{W\,Hz^{-1}}$ and a $\gamma$-ray Eddington ratio $L_{\gamma}/L_{Edd} \gtrsim 0.1$.

In 2021 December, PKS 0735+178 underwent its largest ever recorded flaring event across radio \citep{kadler2021}, optical \citep{zhirkov2021}, X-ray \citep{santander2021} and $\gamma$-ray \citep{garrappa2021} frequencies while being in spacial coincidence with neutrino events reported by \cite{icecube2021}, \cite{dzhilkibaev2021}, \cite{petkov2021} and \cite{filippini2022}. 

We observed PKS 0735+178 over three nights, totalling $11.09\,\mathrm{h}$ in the $griz_\mathrm{s}$ filters. Its \textit{Fermi} data (from the Light Curve Repository (LCR) \citep{abdollahi2023}) showed PKS 0735+178 to be in a $\gamma$-ray state higher than its median flux level over all time, at $4.41\pm0.71\times10^{-7}\,0.1\text{--}100\,\mathrm{GeV\,ph\,cm^{-2}\,s^{-1}}$ (median level: $5.69\pm0.35\times10^{-8}\,0.1\text{--}100\,\mathrm{GeV\,ph\,cm^{-2}\,s^{-1}}$). Its polarimetric properties were measured with MOPTOP, and showed its optical linear polarization degree decreased from roughly 8 to 3 per cent and polarization angle increased from approximately $80^{\circ}$ to $120^{\circ}$ across all $BVR$ bands.

\subsection{OJ248}
OJ248 is a FRSQ at redshift $z\sim0.939$ \citep{massaro2015}. A long-term multiwavelength analysis of this source was performed by the GASP-WEBT Collaboration from 2006--2013 \citep{carnerero2015} including data at radio, NIR, optical, X-ray and $\gamma$-ray frequencies. A large multiwavelength flare was observed in late 2012 by WEBT, \textit{Swift} \citep{dammando2012} and \textit{Fermi} \citep{orienti2012b} as well as an additional radio outburst in late 2010 and optical-NIR flare in early 2007. 

We observed OJ248 over three nights, totalling $8.68\,\mathrm{h}$ in the $griz_\mathrm{s}$ filters. Its optical polarization degree, measured by MOPTOP, was below 5 per cent in V band. \textit{Fermi} $\gamma$-ray data from the Light Curve Repository (LCR) \citep{abdollahi2023} showed OJ248 to be in a $\gamma$-ray state consistent with its median flux level over all time, at $7.02\pm5.44\times10^{-8}\,0.1\text{--}100\,\mathrm{GeV\,ph\,cm^{-2}\,s^{-1}}$ (median level: $8.22\pm0.88\times10^{-8}\,0.1\text{--}100\,\mathrm{GeV\,ph\,cm^{-2}\,s^{-1}}$).

\section{Temporal Variability}
\label{sec: analysis}
The $griz_\mathrm{s}$ light curves for each source on a given night are shown in Figs \ref{txs0506+056_lc}, \ref{oj287_lc}, \ref{pks0735+178_lc} and \ref{oj248_lc} for TXS 0506+056, OJ287, PKS 0735+178 and OJ248 respectively. The source name and night of observation are given above each plot. 

In order to quantify variability in the light curves and to disentangle intrinsic variability from noise, we employ several statistical tests to determine the variability likelihood. Specifically, we calculate the variability amplitude and fractional variability and perform chi-squared and enhanced F-test analyses. These tests are detailed in Sections \ref{sec: var amp}, \ref{sec: frac var}, \ref{sec: chi2}, and \ref{sec: f-test}.

\subsection{Variability Amplitude}
\label{sec: var amp}
Variability amplitude is defined in \cite{heidt1996} as
\begin{equation}
    \mathrm{VA} = \sqrt{(x_{\mathrm{max}}-x_{\mathrm{min}})^2-2\langle x_{\mathrm{err}}\rangle^2}
    \label{var_amp}
\end{equation}
where $x_{\mathrm{max}}$ and $x_{\mathrm{min}}$ are the maximum and minimum observed values, and $\langle x_{\mathrm{err}}\rangle$ is the median measurement error. The percentage variability amplitude, $\mathrm{VA_{per}}$ is given in \cite{romero1999} as 
\begin{equation}
    \mathrm{VA_{per}} = \frac{100}{\langle x\rangle}\sqrt{(x_{\mathrm{max}}-x_{\mathrm{min}})^2-2\langle x_{\mathrm{err}}\rangle^2}
    \label{per_var_amp}
\end{equation}
where $\langle x\rangle$ is the average observed value. Its error, $\Delta \mathrm{VA_{per}}$, is given in \cite{singh2018} as 
\begin{multline}
    \Delta \mathrm{VA_{per}} = 100\times \left(\frac{x_\mathrm{max}-x_\mathrm{min}}{\langle x\rangle\times\mathrm{VA}}\right)\times\\ 
    \sqrt{\left(\frac{x_\mathrm{err,max}}{\langle x\rangle}\right)^2+\left(\frac{x_\mathrm{err,min}}{\langle x\rangle}\right)^2+\left(\frac{\langle x_{err}\rangle}{x_{\mathrm{max}}-x_{\mathrm{min}}}\right)^2 \mathrm{VA}^4}
\end{multline}
where $x_{\mathrm{err,max}}$ and $x_{\mathrm{err,min}}$ are the errors on the maximum and minimum values, respectively. The variability amplitude aims to provide a quantification of the absolute range of variability of a given source by simply looking at the range of magnitudes outside of the scatter from the measurement errors.

The values obtained via the variability amplitude calculations can be seen in column four in Table \ref{var_stats}. We find variability amplitudes ranging from 0.167 per cent up to 1.456 per cent across all sources, dates and filters. The ratio between the error and percentage variability amplitude shows that in 12/48 cases the percentage variability amplitudes are associated with large errors (where the ratio is greater than 3). We note that nine of these cases are attributed to the source OJ248; likely due to it being the faintest of the sample.

\subsection{Fractional Variability}
\label{sec: frac var}
Fractional variability, described fully in \cite{schleicher2019}, is a method of quantifying variability intensity while accounting for measurement uncertainties. It differs from the variability amplitude by considering the variability relative to the mean brightness level. It is defined in \cite{edelson2002} as 
\begin{equation}
    F_\mathrm{var} = \sqrt{\sigma^2_{\mathrm{NXS}}} = \sqrt{\frac{S^2-\langle\sigma^2_{\mathrm{err}}\rangle}{\langle x \rangle^2}}
\end{equation}
where $S^2$ is the variance of the data set, $\langle\sigma^2_{\mathrm{err}}\rangle$ is the median square error, and $\langle x \rangle$ is the median value. It can also be given as the square root of the normalised excess variance ($\sigma^2_{\mathrm{NXS}}$). Its associated error is given in \cite{poutanen2008} as
\begin{equation}
    \Delta F_\mathrm{var} = \sqrt{\mathrm{F^2_{var}}+\Delta\sigma^2_{\mathrm{NXS}}}-F_\mathrm{var}
\end{equation}
where $\Delta\sigma^2_{NXS}$ is the error on the normalised excess variance. This error is given in \cite{vaughan2003} as
\begin{equation}
    \Delta\sigma^2_{\mathrm{NXS}} = \sqrt{\left(\sqrt{\frac{2}{\mathrm{N}}}\frac{\langle\sigma^2_{\mathrm{err}}\rangle}{\langle x \rangle^2}\right)+\left(\sqrt{\frac{\langle\sigma^2_{\mathrm{err}}\rangle}{\mathrm{N}}}\frac{2F_\mathrm{var}}{\langle x\rangle} \right)^2}
\end{equation}
where N is the number of data points in the sample. It follows that if the variance is less than the average square error, $S^2<\langle\sigma^2_{\mathrm{err}}\rangle$, a real value cannot be computed and will be denoted as $<0$, indicating detection of insignificant variability. Where sources had $F_\mathrm{var} > \Delta F_\mathrm{var}$, this test is deemed to show that an object has shown significant variability. 

The fractional variability values are shown in column five in Table \ref{var_stats}. We find 12/48 instances across all sources, dates and filters where significant levels of variability have been detected. These detections correspond to OJ287 on 2023 January 15, and PKS 0735+178 on 2023 January 15 and 17 across all filters, with the most significant detections corresponding to PKS 0735+178 on 2023 January 17.

\subsection{Chi-squared}
\label{sec: chi2}
Chi-squared ($\chi^2$), as used in \cite{zeng2017}, is given by
\begin{equation}
    \chi^2 = \sum_i^N\left(\frac{x_i-\langle x\rangle}{x_{\mathrm{err},i}}\right)^2
\end{equation}
where $x_i$ and $x_{\mathrm{err},i}$ are the individual values and errors respectively within the data set, and $\langle x\rangle$ is the median value. Its critical value was determined at the 99.9 per cent confidence level ($\alpha$ = 0.001) with the degrees of freedom being equal to the number of data points. Where the value is greater than the critical value, significant variability has been detected. $\chi^2$ is a useful metric as it quantifies the levels of variability about the median values. Incorporating the critical value allows us to determine the significance of the value. 

The $\chi^2$ values together with critical values are shown in column six in Table \ref{var_stats}. With the $\chi^2$ test, we detect significant variability in 27/48 instances across all sources, dates and filters. In most cases, non-detection is consistent across filters per source per date. We find consistent variability detections across all filters for TXS 0506+056 on 2023 January 15, OJ287 on 2023 January 15 and 18, and PKS 0735+178 on 2023 January 15 and 17.

\subsection{Enhanced F-test}
\label{sec: f-test}
The enhanced F-test ($F_\mathrm{enh}$) is given in \cite{diego2014} and aims to quantify the variability of a target while incorporating the variability of multiple reference stars \citep{pandian2022}. It is given as
\begin{equation}
    F_{\mathrm{enh}} = \frac{S^2_{\mathrm{blazar}}}{S^2_{\mathrm{star}}}
    \label{ftest}
\end{equation}
where $S^2_{\mathrm{blazar}}$ is the variance of the blazar and $S^2_{\mathrm{star}}$ is the combined variance of the comparison stars. $S^2_{\mathrm{star}}$ is defined as
\begin{equation}
    S^2_{\mathrm{star}} = \frac{1}{(\sum_{j=1}^kN_j-k)}\sum_{j=1}^k\sum_{i=1}^{N_j}\sigma_{j,i}
    \label{star_var}
\end{equation}
where $N_j$ is the number of observations of the $j^\mathrm{th}$ comparison star, k is the number of comparison stars and $\sigma_{j,i}$ is the scaled square deviation. $\sigma_{j,i}$ is given as
\begin{equation}
    \sigma_{j,i} = s_j(m_{j,i}-\langle m_j\rangle)^2
    \label{sigma}
\end{equation}
where $m_{j,i}$ is the magnitudes of each comparison star, $\langle m_j\rangle$ is the mean magnitude of the comparison star and $s_j$ is the scaling factor to account for the different SNRs of the comparison stars. $s_j$ is given as 
\begin{equation}
    s_j = \frac{\langle\sigma_{\mathrm{blazar}}^2\rangle}{\langle\sigma_{sj}^2\rangle}
    \label{s}
\end{equation}
where $\langle\sigma_{\mathrm{blazar}}^2\rangle$ and $\langle\sigma_{sj}^2\rangle$ are the average square errors of the blazar and comparison star magnitudes respectively. Its critical value was determined at the 99.9 per cent confidence level ($\alpha$ = 0.001) with the degrees of freedom for the blazars being one less than the number of observations, and the degrees of freedom for the comparison stars being the sum of one less than the number of observations for each comparison star. 

Finally, the ${F_\mathrm{enh}}$ values with critical values are shown in column seven in Table \ref{var_stats}. They show significant detections of variability in 15/48 instances across all sources, dates and filters. 12 of these detections correspond to observations in all filters of OJ287 on 2023 January 15, and PKS 0735+178 on 2023 January 15 and 17.

\subsection{Temporal Variability Summary}
\begin{figure*}
\centering
\includegraphics[width=\textwidth,height=0.3\textheight]{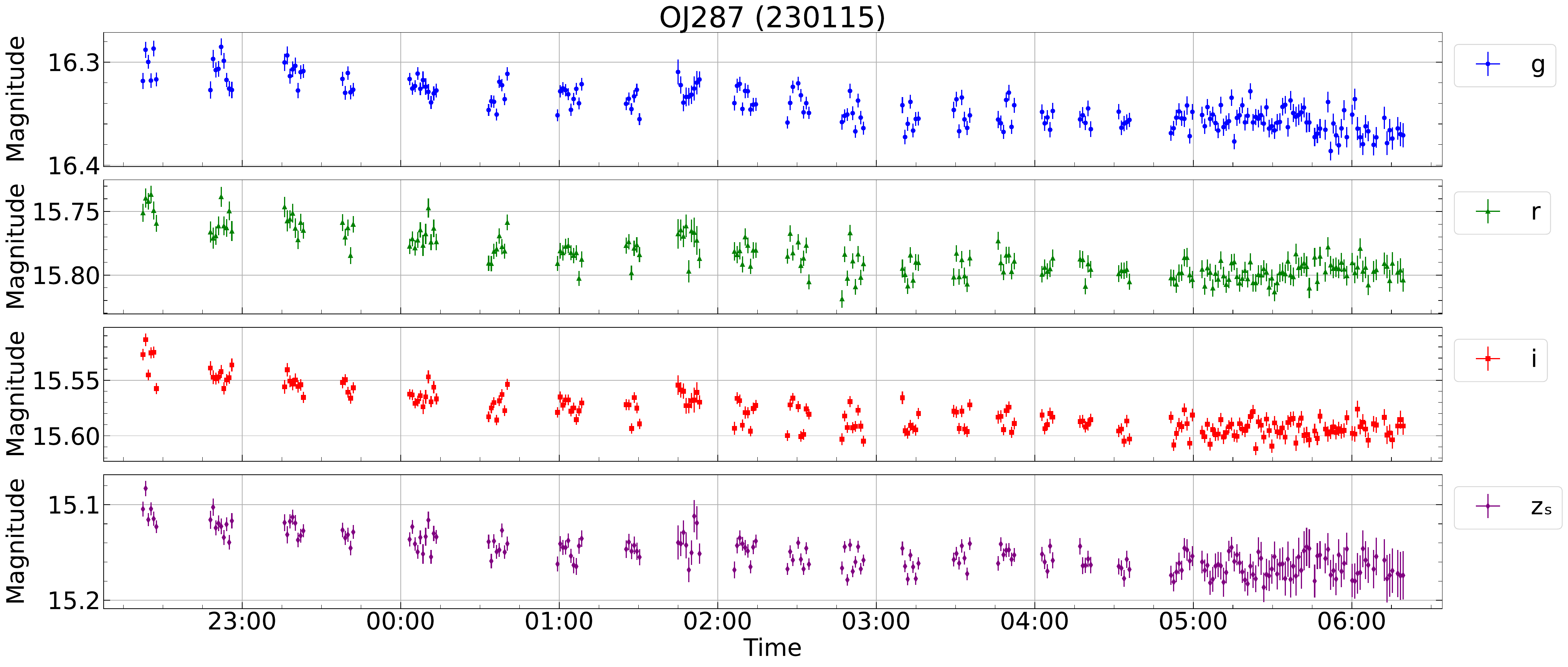}   
\caption{Light curves for OJ287 on the night of 2023 January 15. Panels correspond to $g,r,i,z_\mathrm{s}$ data separately, from top to bottom.}
\label{oj287_lc_15}
\end{figure*}

\begin{figure*}
\centering
\includegraphics[width=\textwidth,height=0.3\textheight]{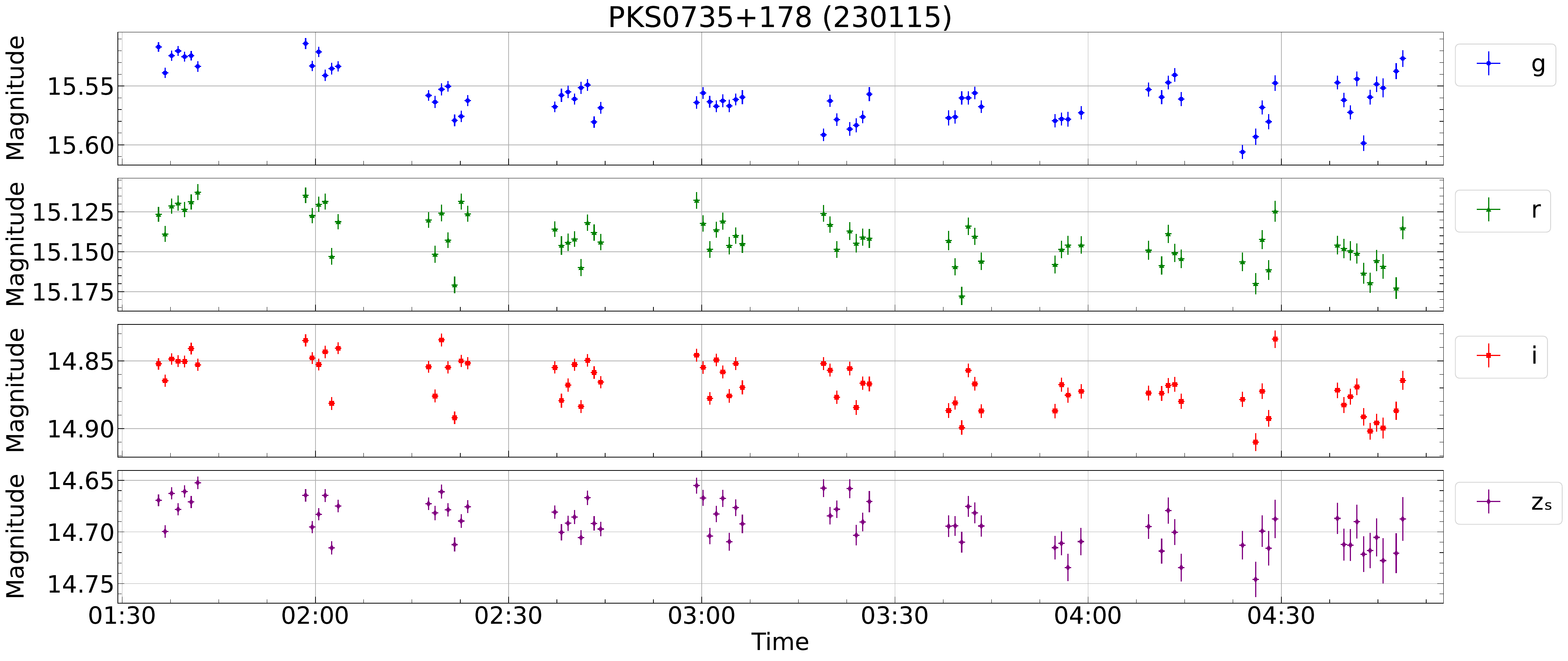}   
\caption{As Fig. \ref{oj287_lc_15}, but for PKS 0735+178 on the night of 2023 January 15.}
\label{pks_lc_15}
\end{figure*}

\begin{figure*}
\centering
\includegraphics[width=\textwidth,height=0.3\textheight]{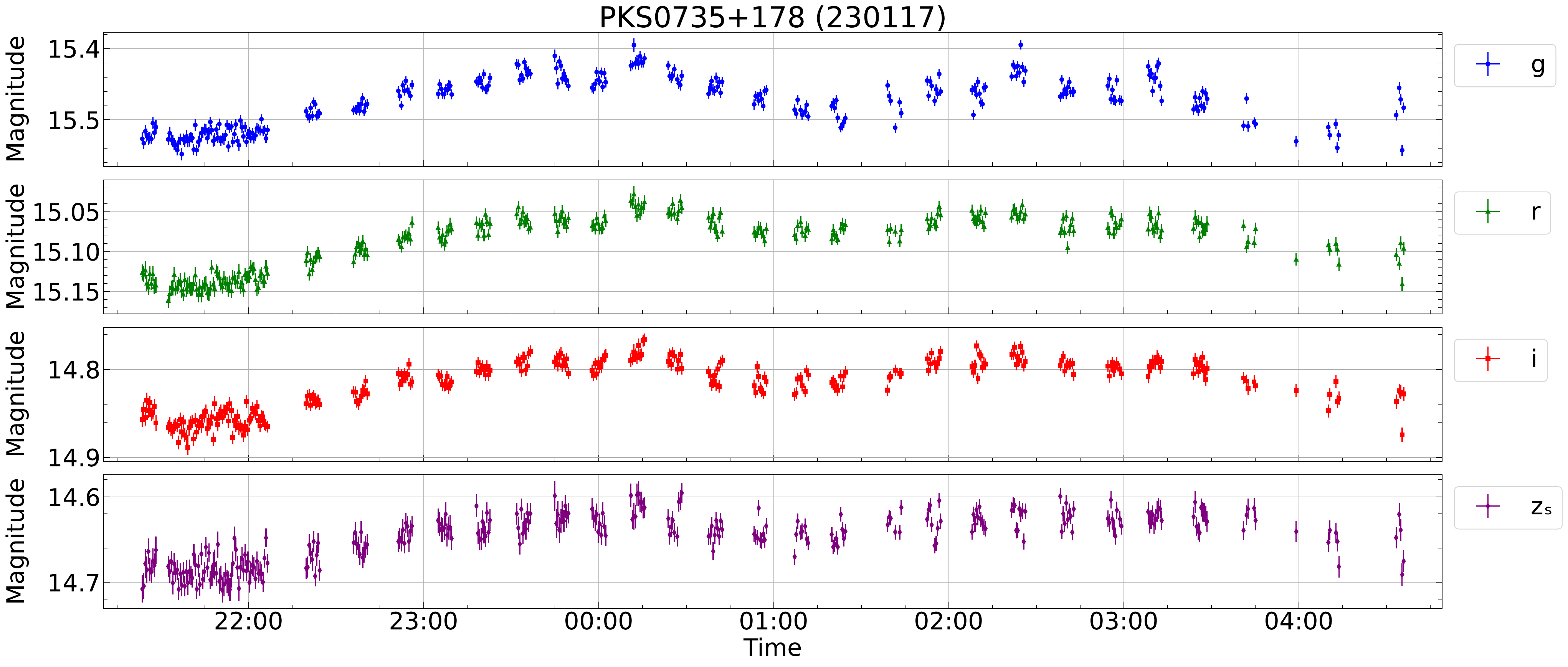}   
\caption{As Fig. \ref{oj287_lc_15}, but for PKS 0735+178 on the night of 2023 January 17.}
\label{pks_lc_17}
\end{figure*}

We summarise the results of each test in Table \ref{var_stats} and determine epochs of variability, we look at the results of the fractional variability, $\chi^2$, and enhanced F-test analyses and determine variable epochs if all three tests are passed. If one or fewer tests were met, we considered there to be no evidence for intranight variability in these sources/epochs.  Overall all TXS 0506+056 and OJ248 epochs display no significant levels of intranight variability. There is possible variability from TXS 0506+056 on 2023 January 16. Given that this possible variability only occurs in the $i$ band, we consider this likely a result of scatter in the data. OJ287 shows one epoch of significant variability on 2023 January 15. PKS 0735+178 shows two epochs of significant variability on 2023 January 15 and 17 and one of possible significant variability on 2023 January 16. The latter consists of very few observations due to poor observing conditions so, based on the light curve, it is possible that this is a false detection. We show the light curves with statistically significant variability for OJ287 and PKS 0735+178 in Figs \ref{oj287_lc_15}, \ref{pks_lc_15}, and \ref{pks_lc_17} and will discuss these variable epochs further in later sections.

\begin{table*}
\centering
\resizebox{0.96\textwidth}{!}{
\begin{tabular}{cccccccc}
\hline
Source       & Date   & Filter & $\mathrm{VA} \pm \Delta \mathrm{VA} (\%) $& $F_\mathrm{var} \pm \Delta F_\mathrm{var}$ & $\chi^2 (\chi^2_{\mathrm{crit}})$ & $F_\mathrm{enh} (F_\mathrm{crit})$ & Variable? \\
\hline
TXS 0506+056 & 230115 & $g$            & 0.311 $\pm$ 0.003 & 0.04 $\pm$ 0.10 & 487.23 (181.99)  & 0.67 (1.54)  & NV \\
...          & ...    & $r$            & 0.368 $\pm$ 0.004 & 0.05 $\pm$ 0.11 & 420.57 (181.99)  & 0.91 (1.54)  & NV \\
...          & ...    & $i$            & 0.398 $\pm$ 0.003 & 0.05 $\pm$ 0.09 & 575.73 (181.99)  & 1.02 (1.54)  & NV \\
...          & ...    & $z_\mathrm{z}$ & 0.499 $\pm$ 0.006 & 0.06 $\pm$ 0.12 & 303.86 (181.99)  & 1.23 (1.54)  & NV \\
TXS 0506+056 & 230116 & g              & 0.225 $\pm$ 0.004 & 0.02 $\pm$ 0.30 & 195.04 (213.97)  & 1.17 (1.48)  & NV \\
...          & ...    & $r$            & 0.291 $\pm$ 0.005 & 0.02 $\pm$ 0.33 & 185.92 (213.97)  & 1.05 (1.48)  & NV \\
...          & ...    & $i$            & 0.258 $\pm$ 0.004 & 0.03 $\pm$ 0.20 & 255.78 (213.97)  & 1.48 (1.48)  & PV \\
...          & ...    & $z_\mathrm{z}$ & 0.294 $\pm$ 0.008 & <0   & 138.42 (213.97)  & 1.13 (1.48)  & NV \\
TXS 0506+056 & 230118 & $g$            & 0.231 $\pm$ 0.004 & 0.03 $\pm$ 0.24 & 151.25 (147.01)  & 0.65 (1.63)  & NV \\
...          & ...    & $r$            & 0.183 $\pm$ 0.004 & <0   & 87.88 (147.01)   & 0.93 (1.63)  & NV \\
...          & ...    & $i$            & 0.167 $\pm$ 0.004 & 0.01 $\pm$ 0.67 & 100.20 (147.01)  & 0.59 (1.63)  & NV \\
...          & ...    & $z_\mathrm{z}$ & 0.221 $\pm$ 0.006 & <0   & 95.45 (147.01)   & 1.34 (1.63)  & NV \\
\hline
OJ287        & 230115 & $g$            & 0.615 $\pm$ 0.006 & 0.12 $\pm$ 0.06 & 1575.69 (281.37) & 1.45 (1.40)  & V  \\
...          & ...    & $r$            & 0.517 $\pm$ 0.005 & 0.09 $\pm$ 0.07 & 1212.16 (281.37) & 1.74 (1.40)  & V  \\
...          & ...    & $i$            & 0.628 $\pm$ 0.005 & 0.11 $\pm$ 0.05 & 2472.15 (281.37) & 1.82 (1.40)  & V  \\
...          & ...    & $z_\mathrm{z}$ & 0.679 $\pm$ 0.010 & 0.11 $\pm$ 0.08 & 962.74 (281.37)  & 1.59 (1.40)  & V  \\
OJ287        & 230118 & $g$            & 0.539 $\pm$ 0.008 & 0.07 $\pm$ 0.13 & 569.87 (369.03)  & 0.94 (1.34)  & NV \\
...          & ...    & $r$            & 0.519 $\pm$ 0.009 & 0.06 $\pm$ 0.15 & 489.41 (369.03)  & 1.04 (1.34)  & NV \\
...          & ...    & $i$            & 0.503 $\pm$ 0.009 & 0.08 $\pm$ 0.10 & 796.89 (369.03)  & 1.23 (1.34)  & NV \\
...          & ...    & $z_\mathrm{z}$ & 0.534 $\pm$ 0.014 & 0.06 $\pm$ 0.16 & 445.58 (369.03)  & 0.86 (1.34)  & NV \\
OJ287        & 230119 & $g$            & 0.398 $\pm$ 0.015 & 0.06 $\pm$ 0.40 & 23.90 (42.31)    & 1.22 (3.07)  & NV \\
...          & ...    & $r$            & 0.497 $\pm$ 0.012 & 0.08 $\pm$ 0.29 & 25.79 (42.31)    & 0.66 (3.07)  & NV \\
...          & ...    & $i$            & 0.189 $\pm$ 0.007 & <0   & 11.46 (42.31)    & 0.26 (3.07)  & NV \\
...          & ...    & $z_\mathrm{z}$ & 0.349 $\pm$ 0.018 & <0   & 7.76 (42.31)     & 0.56 (3.07)  & NV \\
\hline
PKS0735+178  & 230115 & $g$            & 0.591 $\pm$ 0.004 & 0.12 $\pm$ 0.06 & 1194.61 (116.09) & 6.17 (1.76)  & V  \\
...          & ...    & $r$            & 0.427 $\pm$ 0.004 & 0.09 $\pm$ 0.08 & 564.00 (116.09)  & 4.01 (1.76)  & V  \\
...          & ...    & $i$            & 0.509 $\pm$ 0.005 & 0.11 $\pm$ 0.06 & 833.54 (116.09)  & 5.54 (1.76)  & V  \\
...          & ...    & $z_\mathrm{z}$ & 0.632 $\pm$ 0.010 & 0.13 $\pm$ 0.09 & 405.05 (116.09)  & 5.35 (1.76)  & V  \\
PKS0735+178  & 230116 & $g$            & 0.415 $\pm$ 0.005 & 0.08 $\pm$ 0.13 & 138.44 (52.62)   & 5.63 (2.59)  & PV \\
...          & ...    & $r$            & 0.325 $\pm$ 0.005 & 0.06 $\pm$ 0.22 & 61.50 (52.62)    & 3.88 (2.59)  & PV \\
...          & ...    & $i$            & 0.447 $\pm$ 0.005 & 0.08 $\pm$ 0.14 & 109.87 (52.62)   & 6.12 (2.59)  & PV \\
...          & ...    & $z_\mathrm{z}$ & 0.353 $\pm$ 0.008 & 0.06 $\pm$ 0.28 & 46.23 (52.62)    & 2.45 (2.59)  & NV \\
PKS0735+178  & 230117 & $g$            & 0.992 $\pm$ 0.008 & 0.22 $\pm$ 0.03 & 6815.45 (383.68) & 16.00 (1.33) & V  \\
...          & ...    & $r$            & 0.883 $\pm$ 0.009 & 0.22 $\pm$ 0.03 & 5214.03 (383.68) & 21.84 (1.33) & V  \\
...          & ...    & $i$            & 0.825 $\pm$ 0.008 & 0.19 $\pm$ 0.04 & 4846.07 (383.68) & 16.31 (1.33) & V  \\
...          & ...    & $z_\mathrm{z}$ & 0.769 $\pm$ 0.013 & 0.18 $\pm$ 0.07 & 1711.95 (383.68) & 8.24 (1.33)  & V  \\
\hline
OJ248        & 230116 & $g$            & 0.717 $\pm$ 0.020 & 0.08 $\pm$ 0.34 & 27.81 (56.89)    & 0.29 (2.46)  & NV \\
...          & ...    & $r$            & 0.588 $\pm$ 0.020 & 0.04 $\pm$ 0.89 & 27.86 (56.89)    & 0.31 (2.46)  & NV \\
...          & ...    & $i$            & 0.533 $\pm$ 0.027 & <0   & 20.36 (56.89)    & 0.09 (2.46)  & NV \\
...          & ...    & $z_\mathrm{z}$ & 0.946 $\pm$ 0.062 & <0   & 19.79 (56.89)    & 0.43 (2.46)  & NV \\
OJ248        & 230117 & $g$            & 0.746 $\pm$ 0.019 & 0.09 $\pm$ 0.19 & 398.02 (340.74)  & 0.28 (1.35)  & NV \\
...          & ...    & $r$            & 0.886 $\pm$ 0.027 & 0.07 $\pm$ 0.30 & 317.50 (340.74)  & 0.17 (1.35)  & NV \\
...          & ...    & $i$            & 1.193 $\pm$ 0.036 & 0.12 $\pm$ 0.20 & 405.82 (340.74)  & 0.20 (1.35)  & NV \\
...          & ...    & $z_\mathrm{z}$ & 1.456 $\pm$ 0.088 & <0   & 192.61 (340.74)  & 0.22 (1.35)  & NV \\
OJ248        & 230119 & $g$            & 0.775 $\pm$ 0.021 & 0.11 $\pm$ 0.24 & 104.02 (107.26)  & 0.51 (1.81)  & NV \\
...          & ...    & $r$            & 0.967 $\pm$ 0.034 & 0.09 $\pm$ 0.41 & 77.57 (107.26)   & 0.36 (1.81)  & NV \\
...          & ...    & $i$            & 0.884 $\pm$ 0.033 & 0.07 $\pm$ 0.53 & 70.36 (107.26)   & 0.24 (1.81)  & NV \\
...          & ...    & $z_\mathrm{z}$ & 1.135 $\pm$ 0.067 & <0   & 54.00 (107.26)   & 0.56 (1.81)  & NV \\
\hline
\end{tabular}
}
\caption{Variability analysis results. From left to right, the source name, date of observation and filter is given. VA represents the variability amplitude with error as described in section \ref{sec: var amp}, $F_\mathrm{var}$ is the fractional variability with error as described in section \ref{sec: frac var}, $\chi^2$ is the Chi-squared value given with the critical value as described in section \ref{sec: chi2} and $F_\mathrm{enh}$ is the enhanced F-test value with the critical value as described in section \ref{sec: f-test}. The final column describes whether or not the source was deemed variable in a particular filter on a given night. Of the three variability tests ($F_\mathrm{var}$, $\chi^2$ and $F_\mathrm{enh}$), if all three showed variability the source was deemed variable (V), two meant possibly variable (PV), and one or none meant likely not variable (NV).}
\label{var_stats}
\end{table*}

\section{Colour variability}
\label{sec: colour}
We test for colour variability by investigating how the $g-z_\mathrm{s}$ colour changes with respect to $r$ band magnitude and with time. We use the Spearman rank correlation coefficients to quantify the level of monotonic variability observed in each source on a given night. We use $\alpha=0.05$ for the significance coefficient, p, implying a 95 per cent confidence interval and assign the strength of the correlation, c, by the ranges specified in Table \ref{srank}. We also utilise the enhanced F-test to account for the variability of the reference stars as previously described. 

\begin{table}
\centering
\begin{tabular}{cc}\\ \hline
Value & Correlation Degree\\ \hline
$\textit{c}$ = 0 & no correlation \\
0 $\leq$ |$\textit{c}$| < 0.2 & very weak \\
0.2 $\leq$ |$\textit{c}$| < 0.4 & weak \\
0.4 $\leq$ |$\textit{c}$| < 0.6 & moderate \\
0.6 $\leq$ |$\textit{c}$| < 0.8 & strong \\
0.8 $\leq$ |$\textit{c}$| < 1 & very strong \\
|$\textit{c}$| = 1 & monotonic \\
\hline
\end{tabular}
\caption{Correlation strengths for Spearman rank correlation coefficients. The magnitude of c shows whether the correlation is positive or negative.}
\label{srank}
\end{table}

The Spearman rank correlation coefficients and enhanced F-test values for each set of $g-z_\mathrm{s}$ vs $r$ data are shown in Table \ref{colour_stats} (full table for all epochs available in Table \ref{colour_stats_full} with the corresponding plots in Fig. \ref{colour plots} in the appendix). There are two epochs that show significant colour variability during observations; PKS 0735+178 on the nights of 2023 January 15 and 17. The former shows a positive correlation with a strength of 0.33, indicating a weak correlation. The positive nature of this correlation implies as the source gets brighter, it also gets redder in colour. Conversely, the latter date shows a negative correlation with  a strength of 0.35, again indicating a weak correlation, although the negative nature this time indicates as the source gets brighter it also gets bluer. 

To confirm these results, we also obtained correlation statistics on the slope of the optical spectral energy distribution (SED) vs the $r$ band magnitude. The slope was obtained by fitting a line through the $g,i,z_\mathrm{s}$ band magnitudes at each epoch. This analysis confirmed the same sources and epochs to show significant variability as the colour analysis (see Appendix \ref{sec: spectral index} for details).

\begin{table*}
\centering
\begin{tabular}{cccccc}
\hline
Source        & Date        & $p$                 & $c$   & $F_\mathrm{enh}$ ($F_\mathrm{crit}$) & Variable? \\
\hline
OJ287         & 2023 Jan 15 & 0.07                & -0.12 & 0.88 (1.53)            & NV        \\

PKS 0735+178  & 2023 Jan 15 & 4.6$\times10^{-3}$  & 0.33  & 4.66 (1.72)            & V         \\
…             & 2023 Jan 16 & 0.7                 & -0.08 & 1.84 (2.40)            & NV        \\
…             & 2023 Jan 17 & 3.9$\times10^{-10}$ & -0.35 & 3.13 (1.31)            & V         \\
\hline
\end{tabular}
\caption{Colour variability statistics for variable sources on a given night. $p$ and $c$ refer to the Spearman rank correlation coefficients (significance and strength respectively), $F_\mathrm{enh}$ is the enhanced F-test value with the critical value as described in section \ref{sec: f-test}, and the final column describes whether or not the colour of the source was deemed variable on the given night. If $p<0.05$ and $F_\mathrm{enh}>F_\mathrm{crit}$ the source was deemed variable (V), otherwise not variable (NV).}
\label{colour_stats}
\end{table*}

\section{Time-lag Analysis}
\label{sec: lags}
We test for the possibility of a time lag between $griz_\mathrm{s}$ bands on the nights where sources show statistically significant variability. This would be indicative of a shock or any energy density evolution within the jet, and allow us to rule out geometric variability processes like Doppler factor evolution of spiralling emitting regions. The variability must occur over time-scales less than the duration of the observations (minima and maxima within the lightcurves), which allows us to match up light curve features between bands and test for intra-band lags. Only one source and night fit these criteria, PKS 0735+178 on 2023 January 17. To perform the lag analysis, we utilise the Discrete Correlation Function (DCF) which provides an estimate for the time lag between two unevenly sampled time series without the need for interpolation, while accounting for the effects of correlated errors \citep{edelson1988}. It is defined by 
\begin{equation}
    UDCF_{ij} = \frac{\left(x_i-\langle x\rangle\right)\left(y_j-\langle y\rangle\right)}{\sqrt{\left(\sigma^2_x-\langle\Delta x\rangle^2\right)\left(\sigma^2_y-\langle\Delta y\rangle^2\right)}},
\end{equation}
where ($x_i,y_j$) are the observations, ($\langle x\rangle,\langle y\rangle$) are the mean value from each light curve, ($\sigma_x,\sigma_y$) are the standard deviation of each light curve, and ($\langle\Delta x\rangle,\langle\Delta y\rangle$) are the median error values \citep{liodakis2018}. To find the DCF value associated with each time shift, $\tau$, we average over the number of ($x_i,y_j$) pairs, $N$, where $\tau - \frac{\Delta\tau}{2} < \Delta t_{ij} < \tau + \frac{\Delta\tau}{2}$ or in this case, the mean $UDCF_{ij}$ value
\begin{equation}
    DCF(\tau) = \frac{\sum{UDCF_{ij}}}{N} = \langle UDCF_{ij}\rangle.
\end{equation}
What also sets the DCF apart from other correlation methods is that a standard error on $DCF(\tau)$ can be given by
\begin{equation}
    \Delta DCF(\tau) = \frac{1}{N-1}\left(\sqrt{\sum\left(UDCF_{ij}-DCF(\tau)\right)^2}\right),
\end{equation}
assuming the individual $\mathrm{UDCF_{ij}}$ values within a bin are uncorrelated.

\begin{figure*}
    \centering
    \includegraphics[width=\linewidth]{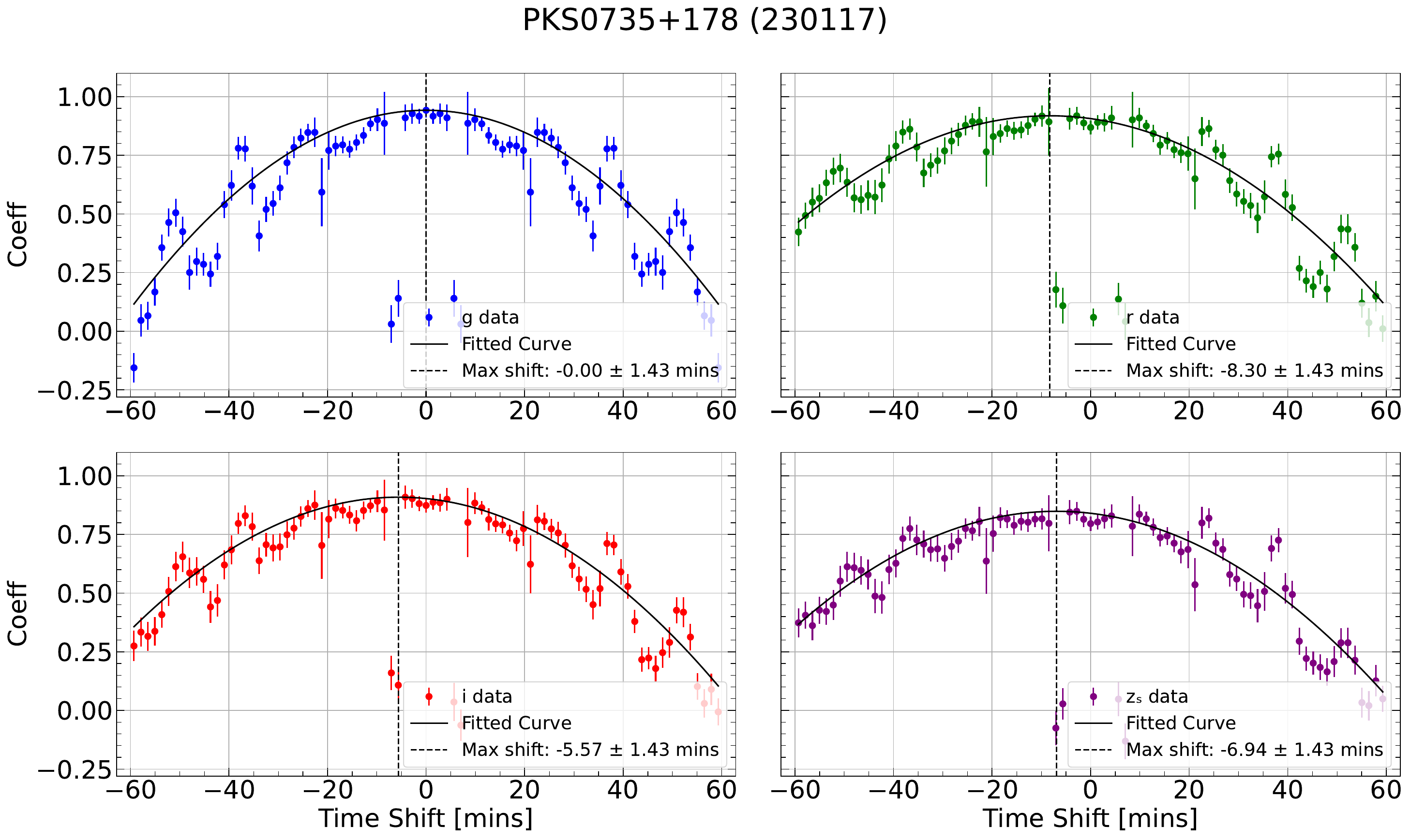}
    \caption{DCF coefficients testing for a lag on the data from PKS 0735+178 on the night of 2023 January 17. The coefficients (blue, green, red, and purple points for filters $g,r,i,z_\mathrm{s}$ respectively) are fitted with a Gaussian (black line) to find the peak. This peak value (vertical dotted line) is shown in the legend with an uncertainty.}
    \label{pks_lag}
\end{figure*}

We investigate the possibility of a lag within $\pm\,60\,\mathrm{min}$. While analysing the data using the DCF, its limitations in accounting for regularly unevenly sampled data became apparent. The data consists of an observing sequence over $\sim5\,\mathrm{min}$ before a $\sim10\,\mathrm{min}$ break whilst observing a second target. When performing the DCF, this periodically resulted in a large decrease in the number of overlapping bins, zero in some instances, within $\frac{\Delta\tau}{2}$. This is seen in the correlation curves (Fig. \ref{pks_lag}) as periodic peaks and drops in the coefficient values. 

Fig. \ref{pks_lag} shows the results of the DCF on the data from PKS 0735+178 on the night of 2023 January 17 on each $griz_\mathrm{s}$ light curve with respect to the $g$ light curve. In this configuration, a positive lag implies $g$ leading the other bands and a negative lag implies $g$ lagging the other bands. The solid curve shows a Gaussian fit to the DCF correlation values, calculated to offset the structure induced by the periodic number of overlapping bins. The dotted line shows the peak of the Gaussian curve, and therefore the lag value. It shows a significant non-zero lag in each $riz_\mathrm{s}$ light curve with respect to $g$. Between the three bands, the lags are all consistent, with a mean value of $-6.94 \pm 1.43\,\mathrm{min}$. The uncertainty of $1.43\,\mathrm{min}$ is the average cadence of the observations and was chosen as the larger value of average cadence and error on the Gaussian peak. 

\begin{figure*}
    \centering
    \includegraphics[width=\linewidth]{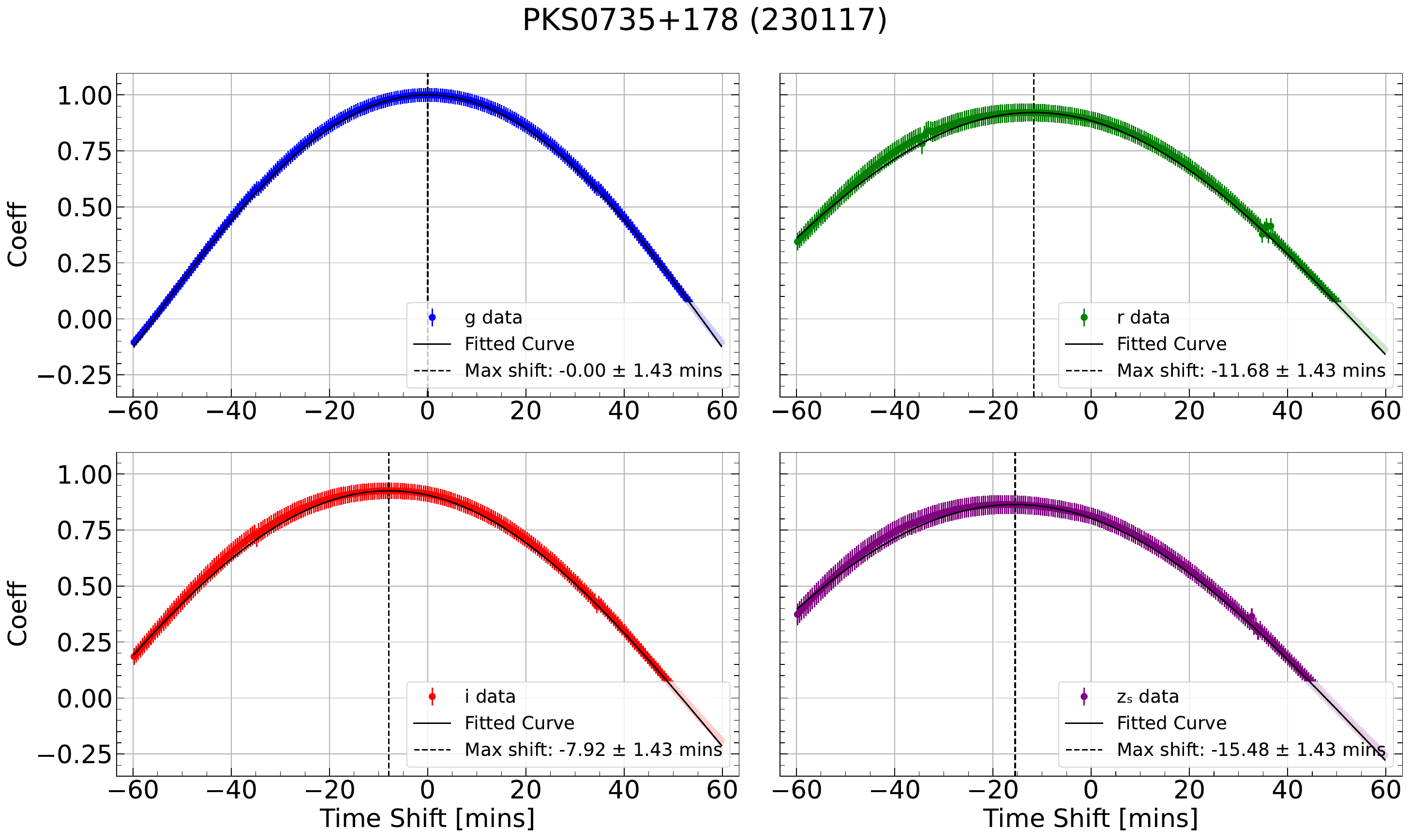}
    \caption{As Fig. \ref{pks_lag} but with the fitted data from PKS 0735+178 on the night of 2023 January 17.}
    \label{pks_lag_fitted}
\end{figure*}

In order to check the significance of the induced correlation curve structure, and to mitigate the scatter in the light curves, we also calculated the DCF after fitting a curve to the data. We fit each light curve using the \textsc{GaussianProcessRegressor} module from \textsc{scikit-learn} in Python \citep{scikit-learn} using the Rational Quadratic kernel. Calculating the DCF on this fitted curve and following the same steps as outlined previously, we obtain the results shown in Fig. \ref{pks_lag_fitted}. We keep the same uncertainties ($1.43\,\mathrm{min}$) to reflect the original data cadence. The results for $g$ and $i$ are consistent with the values obtained previously, but the lags obtained in $r$ and $z_\mathrm{s}$ are significantly larger at $-11.68 \pm 1.43\,\mathrm{min}$ and  $-15.48 \pm 1.43\,\mathrm{min}$, respectively.

\section{Discussion}
\label{sec: discussion}
Blazar intranight variability is thought to arise from geometric changes within the blazar jet; such as the Doppler factor variability of an emitting region travelling in a helical motion in the jet, from the evolution of an emitting region through the jet or from the acceleration/cooling of particles. Additionally, it is entirely possible for the observed behaviour to be a combination of multiple emitting regions or different processes occurring simultaneously.

The mechanism behind Doppler factor variability involves an emitting region, or `blob', of density inhomogeneity travelling helically along the jet. This causes quasi-periodic oscillations (QPOs) in the light curve resulting from the apparent changing Doppler factor and subsequent bulk Lorentz factor \citep{camenzind1992,mohan2015,bachev2023}. On intranight time-scales, this behaviour would present across the optical regime as multiple brightness peaks, depending on the number of blobs, where individual peaks would be observed with no colour changes or time-lags \citep{papadakis2004,bachev2015}. If the origin of the variability was many emitting blobs, each with differing SEDs, then one might expect the emission of different blobs to dominate at different times and subsequently cause rapid colour changes in addition to the brightness changes \citep{bachev2023}. This variability, however, is a relativistic effect rather than any change in the emission output of the source. 

Changes in the intrinsic luminosity of the source on intranight time-scales can be attributed to processes such as shocks or magnetic reconnection in the jet. These processes involve a uniform injection of fresh, more energetic electrons which evolve as a function of their energy distribution, where harder electrons cool faster \citep{urry1997}. This may produce intra- and inter-band time-lags, which can determine cooling times and constrain the homogeneous synchrotron model \citep{tavecchio1998}. An evolving energy distribution may also produce colour variability \citep{papadakis2004}. Additionally, emission at optical frequencies can trace slightly different parts of the SED depending on the location of the synchrotron peak. For LSP sources (three of our sources), optical frequencies trace the falling region of the synchrotron peak which means redder frequencies map higher-energy emission and may produce faster-evolving variability, causing colour variability and time-lags between wavebands. Conversely, for HSP blazars, optical frequencies trace the rising part of the SED so one would expect the bluer frequencies to evolve faster \citep{pandian2022}. 


In our work, we found that TXS 0506+056 and OJ248 showed no evidence of variability in the epochs studied. OJ248 is the faintest object in our sample and would have benefitted from longer exposure times for better signal-to-noise had the autoguider on the TCS been available. TXS 0506+056 showed significant, weak, colour variability on 2023 January 15, which may be due to the scatter in the data. 

OJ287 showed evidence of significant flux variability on the night of 2023 January 15, but no significant changes in colour. There are no significant short-time-scale features in the light curve, and the observed variability consists of a gradual decrease in the brightness over the $\sim$ 6 hours of observing. 

PKS 0735+178 displayed significant variability on two out of three nights, including significant colour correlations showing both redder-when-brighter and bluer-when-brighter behaviour. Additionally, on the night when BWB colour variability was observed, a hard-lag of order $10\,\mathrm{min}$ was detected. 

If the hard-lags observed in PKS 0735+178 are caused by the evolution of the electron energy distribution, different shock-in-jet processes can be examined to explain the variability. When the acceleration time-scale during the shock is much less than the post-shock cooling time-scale, ie $t_\mathrm{acc}\ll t_\mathrm{cool}$, soft lags are expected, where the lower energy emission (red) lag behind the higher energy emission (blue). Conversely, when the acceleration time-scale is comparable to the cooling time-scale, ie. $t_\mathrm{acc}\approx t_\mathrm{cool}$, hard-lags are expected, where the lower energy emission precedes the higher energy emission \citep{zhang2002b}. 

In order to achieve a hard-lag, and produce comparable acceleration and cooling time-scales, an energy injection is required to accelerate electrons after the shock has passed, rather than allowing the shocked particles to cool, which results in soft lags \citep{mastichiadis2008}. Injecting energy into the post-shocked medium can be achieved using second-order Fermi acceleration processes. \cite{kalita2023} describe how turbulent magnetic fields built behind a shock travelling through an inhomogeneous medium can produce these processes, resulting in acceleration of the post-shock particles via magnetic reconnection. In this scenario, energy is released to the surrounding particles through the interaction of magnetic field lines with opposite polarity. 

While we cannot make a firm conclusion as to the nature of the detected INOV in PKS 0735+178, the detection of a hard-lag favours changes to the electron energy distributions and the internal shock model over any geometric changes.

\section{Conclusion}
We performed simultaneous $g,r,i,z_\mathrm{s}$ photometric observations using MuSCAT2 on the Carlos Sánchez Telescope to study the intranight optical variability of four $\gamma$-ray bright blazars. Our analysis consisted of employing several statistical methods to test for the presence of variability on time-scales of a few hours. Additionally, the DCF was used to test for intra-band time lags between bands $r$, $i$, and $z_\mathrm{s}$ with respect to band $g$. We found:
\begin{itemize}
    \item TXS 0506+056 and OJ248 showed no evidence for intranight variability on any night.
    \item OJ287 showed evidence for intranight variability on 2023 January 15. The nature of this variability was a gradual change, around 0.1 magnitudes over 7 hours, and was not accompanied by any significant changes in colour. 
    \item PKS 0735+178 showed evidence for intranight variability on two occasions along with changes in colour; showing both a redder-when-brighter and a bluer-when-brighter correlation on different dates. 
    \item PKS 0735+178 showed a time lag where the $g$ band lags the $r,i,z_\mathrm{s}$ bands by around $10\,\mathrm{min}$. This suggests the variability may arise from changes in the electron energy-density distribution. 
\end{itemize}

Further observations of blazars during all activity states at high cadences can confirm whether intra-band hard-lags across optical frequencies are a more common feature than previously thought. This would provide strong evidence for changes in the jet's energy density as the cause for INOV in blazars.

\section*{Acknowledgements}
We would like to thank the reviewer for their constructive comments.
This article is based on observations made with the Carlos Sanchez Telescope operated on the island of Tenerife by the Instituto de Astrofiscia de Canarias in the Spanish Observatorio del Teide. The Liverpool Telescope is operated on the island of La Palma by Liverpool John Moores University in the Spanish Observatorio del Roque de los Muchachos of the Instituto de Astrofisica de Canarias with financial support from the UKRI Science and Technology Facilities Council (STFC) (ST/T00147X/1).  HEJ acknowledges travel support from the UKRI STFC PATT grant (ST/S001530/1).

\section*{Data Availability}
The data underlying this article will be shared on reasonable request to the corresponding author.



\bibliographystyle{mnras}
\bibliography{paper}



\appendix
\section{Spectral index}
\label{sec: spectral index}
\begin{figure*}
    \centering
    \includegraphics[width=\linewidth]{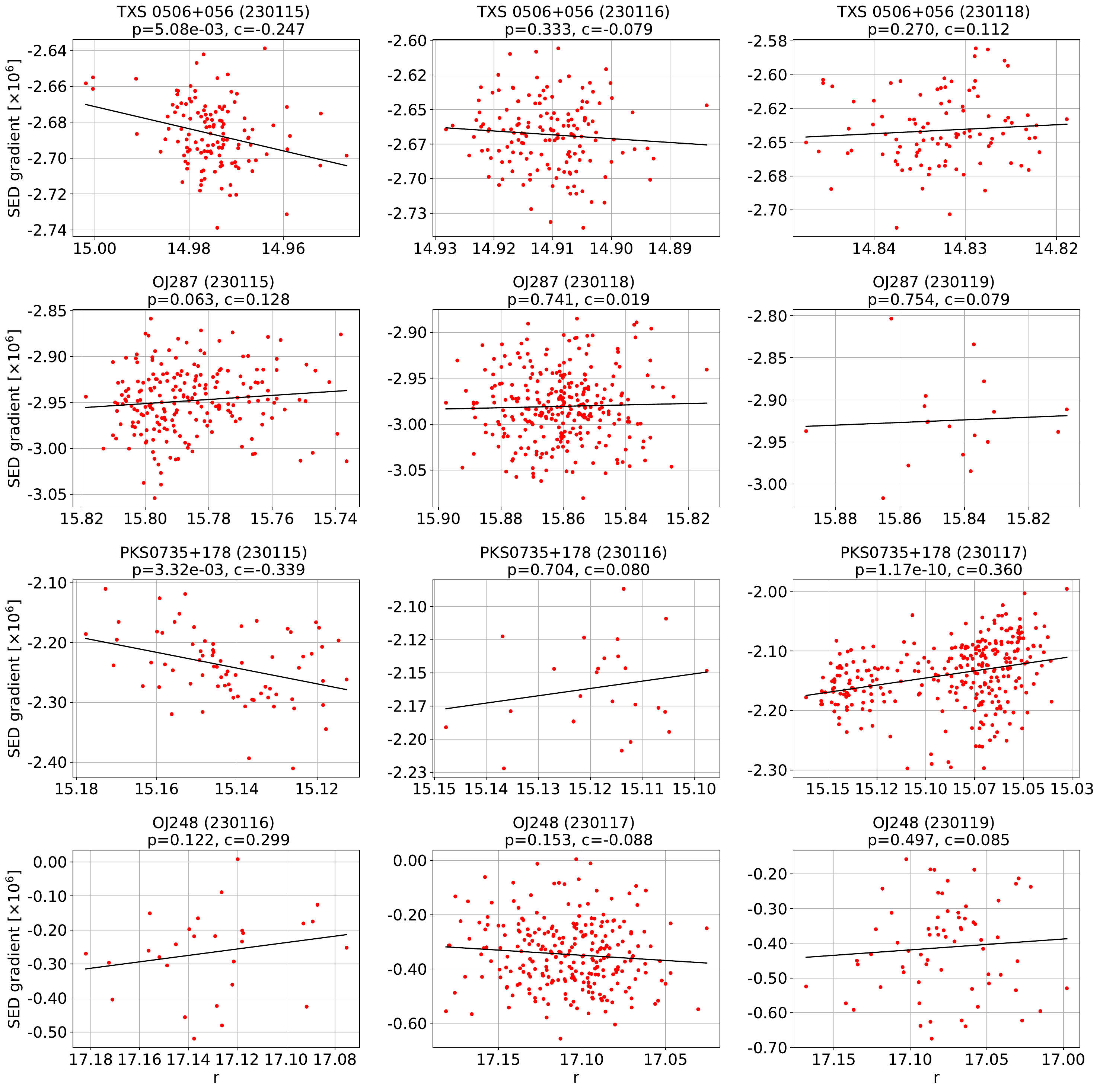}
    \caption{SED gradient using $g$, $i$, and $z_\mathrm{z}$ band magnitudes against $r$ band magnitude for each of the four blazars (different rows) on different nights (different columns) as indicated above each plot. Spearman rank correlation coefficients and significance values are also shown above each plot.}
    \label{sed colour plots}
\end{figure*}

\begin{figure*}
    \centering
    \includegraphics[width=\linewidth]{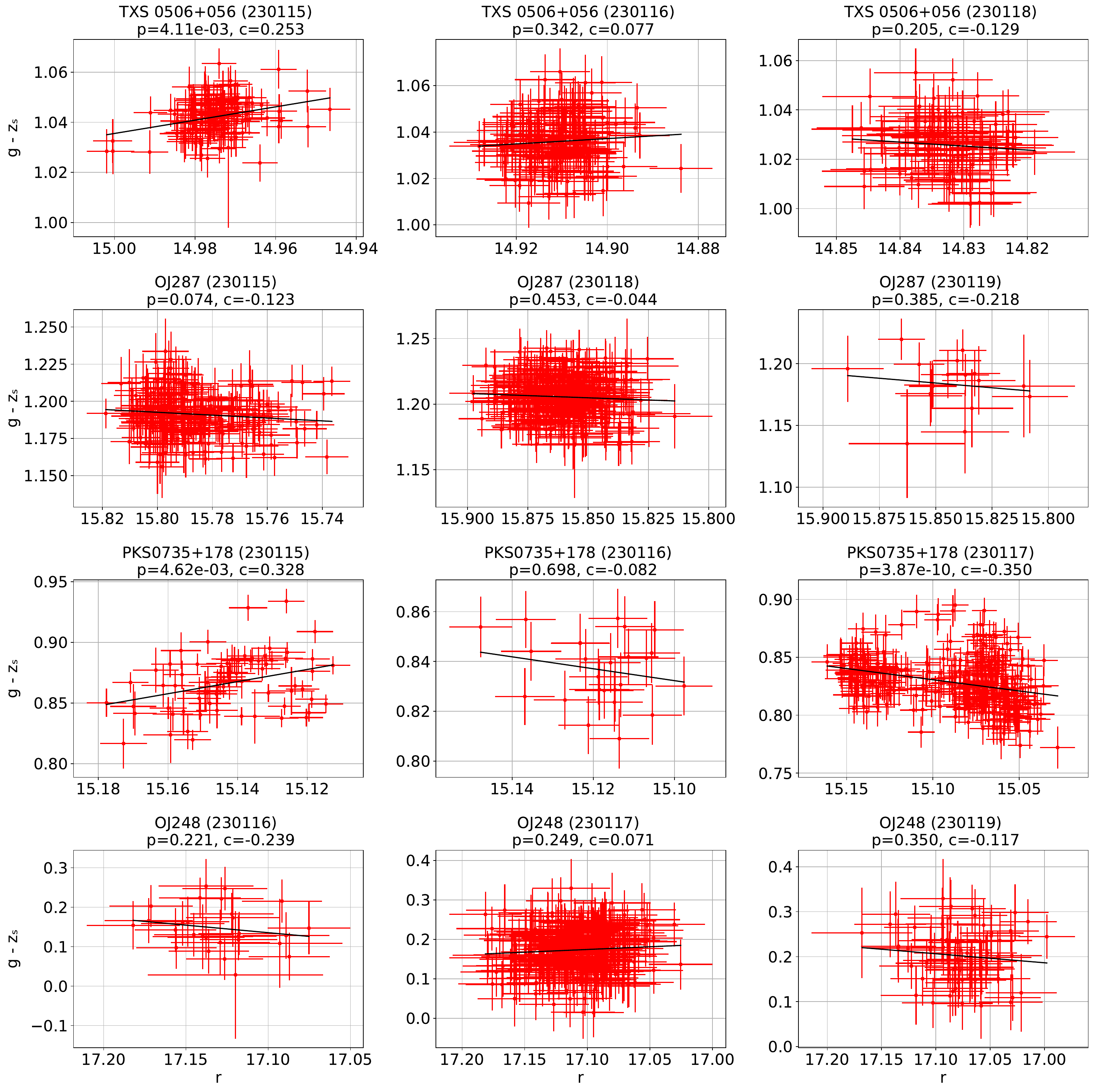}
    \caption{Colour-magnitude ($r$ vs $g$-$z_\mathrm{s}$ magnitudes) diagrams for each of the four blazars (different rows) on different nights (different columns) as indicated above each plot. Also present above each plot are the corresponding Spearman rank correlation coefficients and significance values.}
    \label{colour plots}
\end{figure*}

We study the change in SED of each source on a given night by calculating the gradient of a straight line fitted through the $g$, $i$, and $z_\mathrm{z}$ band magnitudes at each epoch, and correlating it against the corresponding $r$ band magnitude. These plots are shown in Fig. \ref{sed colour plots} where Spearman rank correlation coefficients and significance values are given above each plot. The results align very closely with what is seen in the colour-magnitude diagrams in Fig. \ref{colour plots}, showing the same significance values for each source with very similar correlation strengths. 

\section{Figures and tables}
\begin{figure*}
\centering
\begin{subfigure}{\textwidth}
    \includegraphics[width=\textwidth,height=0.3\textheight]{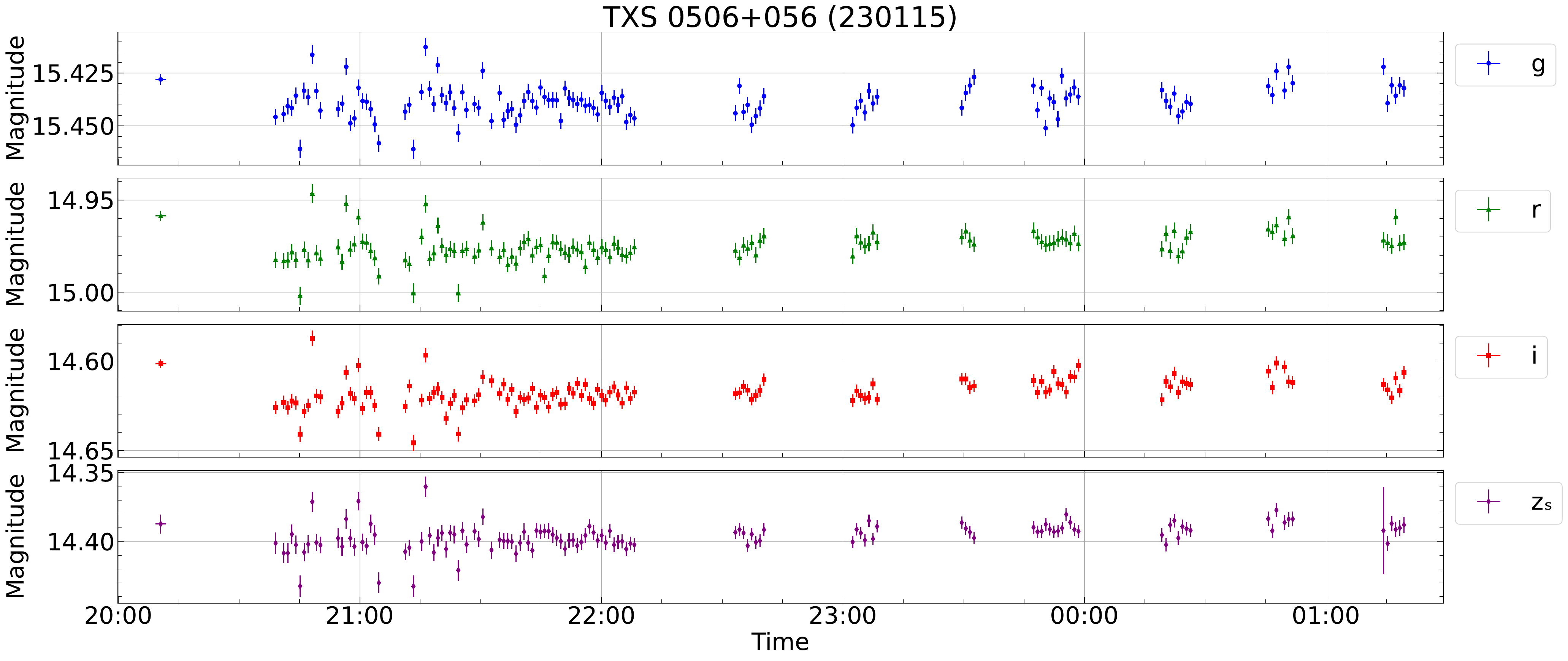}
\end{subfigure}
\hfill
\begin{subfigure}{\textwidth}
    \includegraphics[width=\textwidth,height=0.3\textheight]{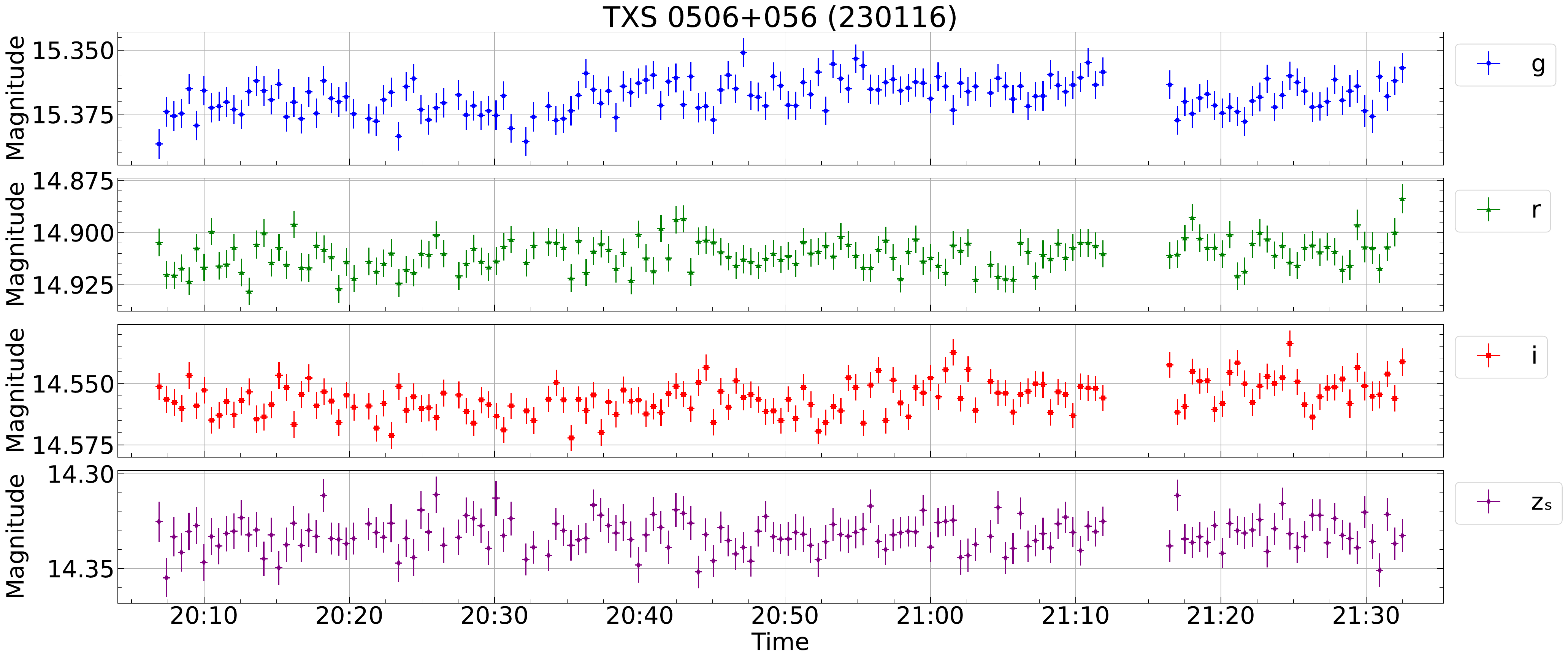}
\end{subfigure}
\hfill
\begin{subfigure}{\textwidth}
    \includegraphics[width=\textwidth,height=0.3\textheight]{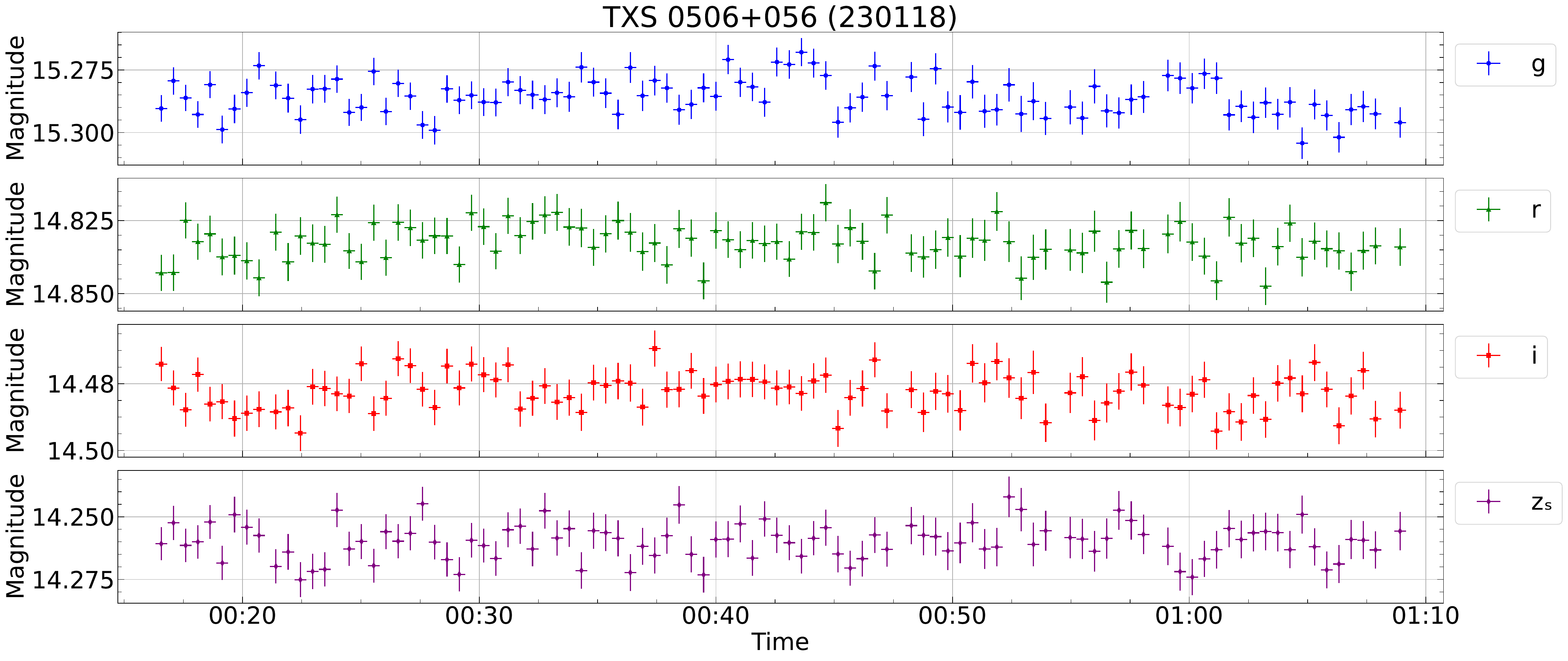}
\end{subfigure}
        
\caption{Light curves of TXS 0506+056 on the nights of 2023 January 15, 2023 January 16, and 2023 January 18. The panels of each of the three plots correspond to $g$, $r$, $i$, and $z_\mathrm{s}$ filters, from top to bottom.}
\label{txs0506+056_lc}
\end{figure*}

\begin{figure*}
\centering
\begin{subfigure}{\textwidth}
    \includegraphics[width=\textwidth,height=0.3\textheight]{light_curves/OJ287_lc_15.pdf}
\end{subfigure}
\hfill
\begin{subfigure}{\textwidth}
    \includegraphics[width=\textwidth,height=0.3\textheight]{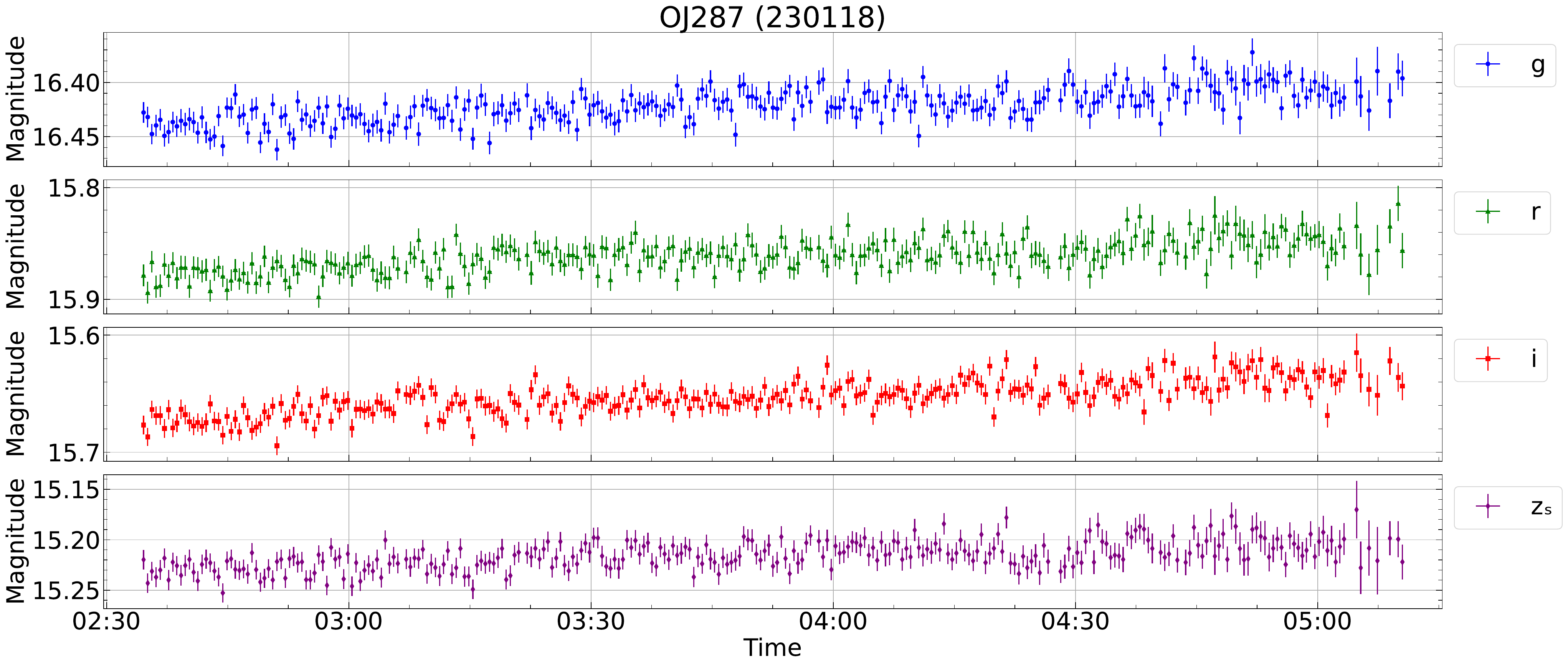}
\end{subfigure}
\hfill
\begin{subfigure}{\textwidth}
    \includegraphics[width=\textwidth,height=0.3\textheight]{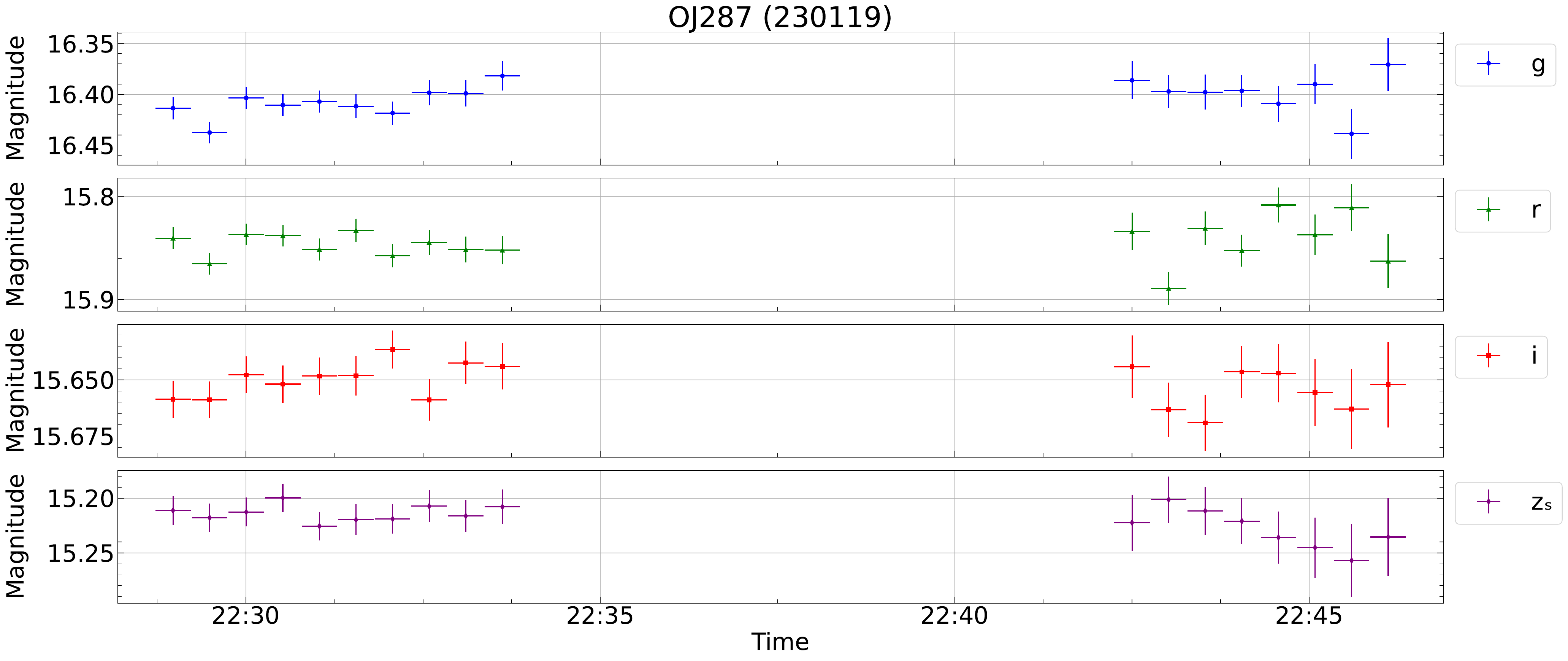}
\end{subfigure}
        
\caption{As Fig. \ref{txs0506+056_lc}, but for OJ287 on the nights of 2023 January 15, 2023 January 18, and 2023 January 19.}
\label{oj287_lc}
\end{figure*}

\begin{figure*}
\centering
\begin{subfigure}{\textwidth}
    \includegraphics[width=\textwidth,height=0.3\textheight]{light_curves/PKS0735+178_lc_15.pdf}
\end{subfigure}
\hfill
\begin{subfigure}{\textwidth}
    \includegraphics[width=\textwidth,height=0.3\textheight]{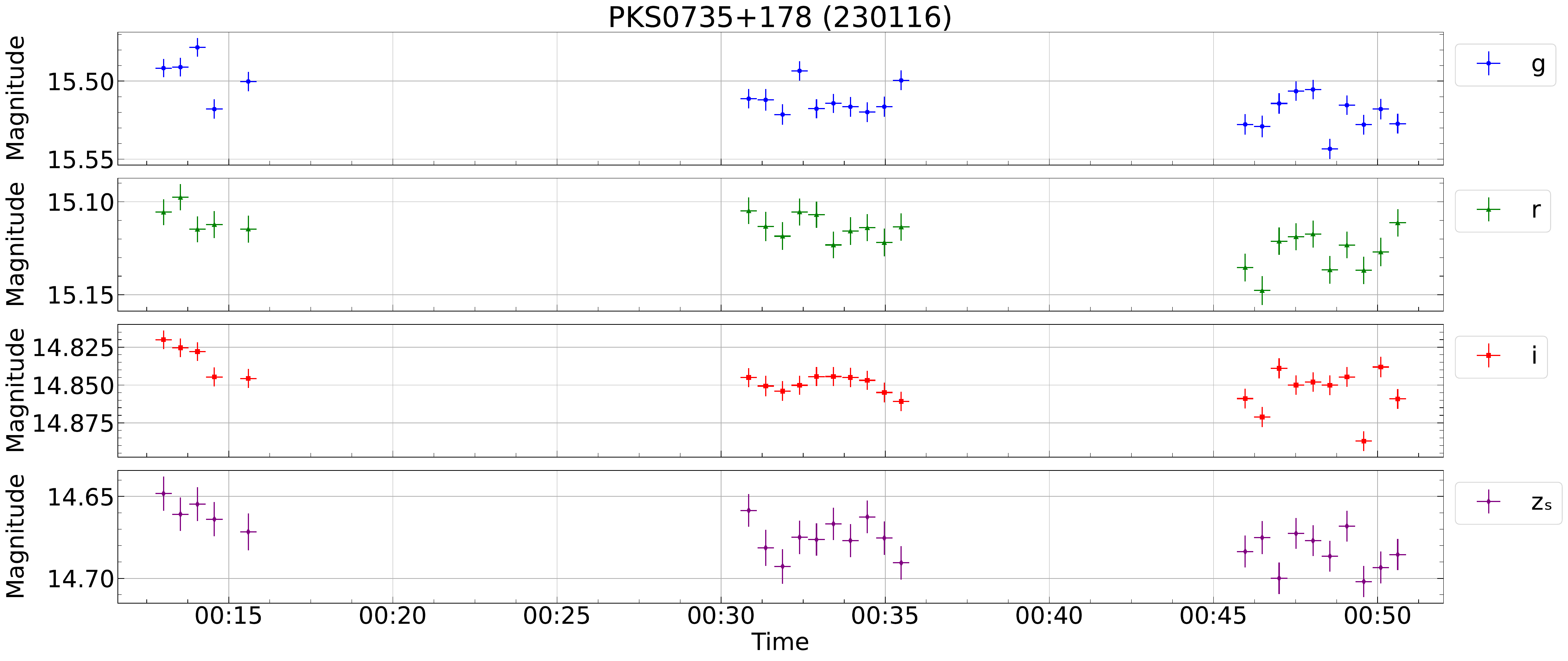}
\end{subfigure}
\hfill
\begin{subfigure}{\textwidth}
    \includegraphics[width=\textwidth,height=0.3\textheight]{light_curves/PKS0735+178_lc_17.pdf}
\end{subfigure}
        
\caption{As Fig. \ref{txs0506+056_lc}, but for PKS 0735+178 on the nights of 2023 January 15, 2023 January 16, and 2023 January 17.}
\label{pks0735+178_lc}
\end{figure*}

\begin{figure*}
\centering
\begin{subfigure}{\textwidth}
    \includegraphics[width=\textwidth,height=0.3\textheight]{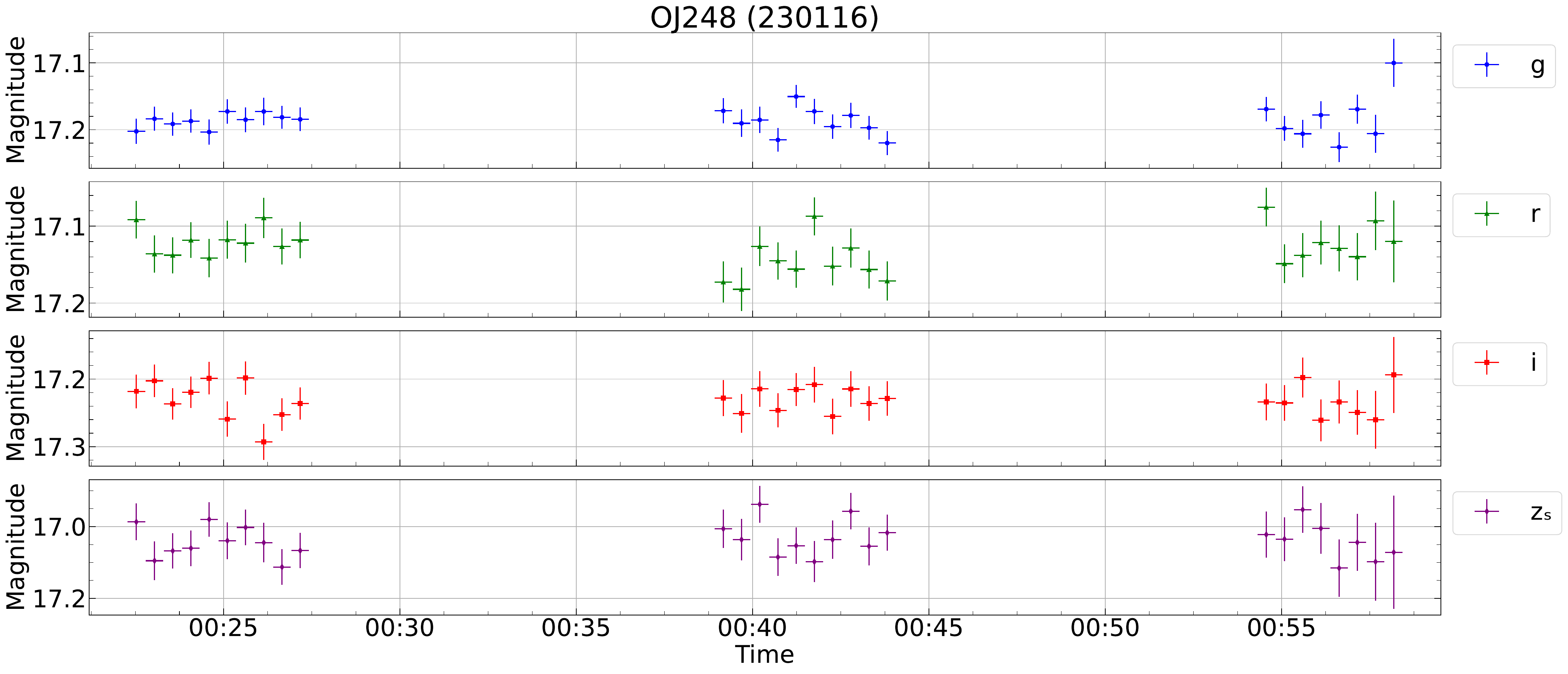}
\end{subfigure}
\hfill
\begin{subfigure}{\textwidth}
    \includegraphics[width=\textwidth,height=0.3\textheight]{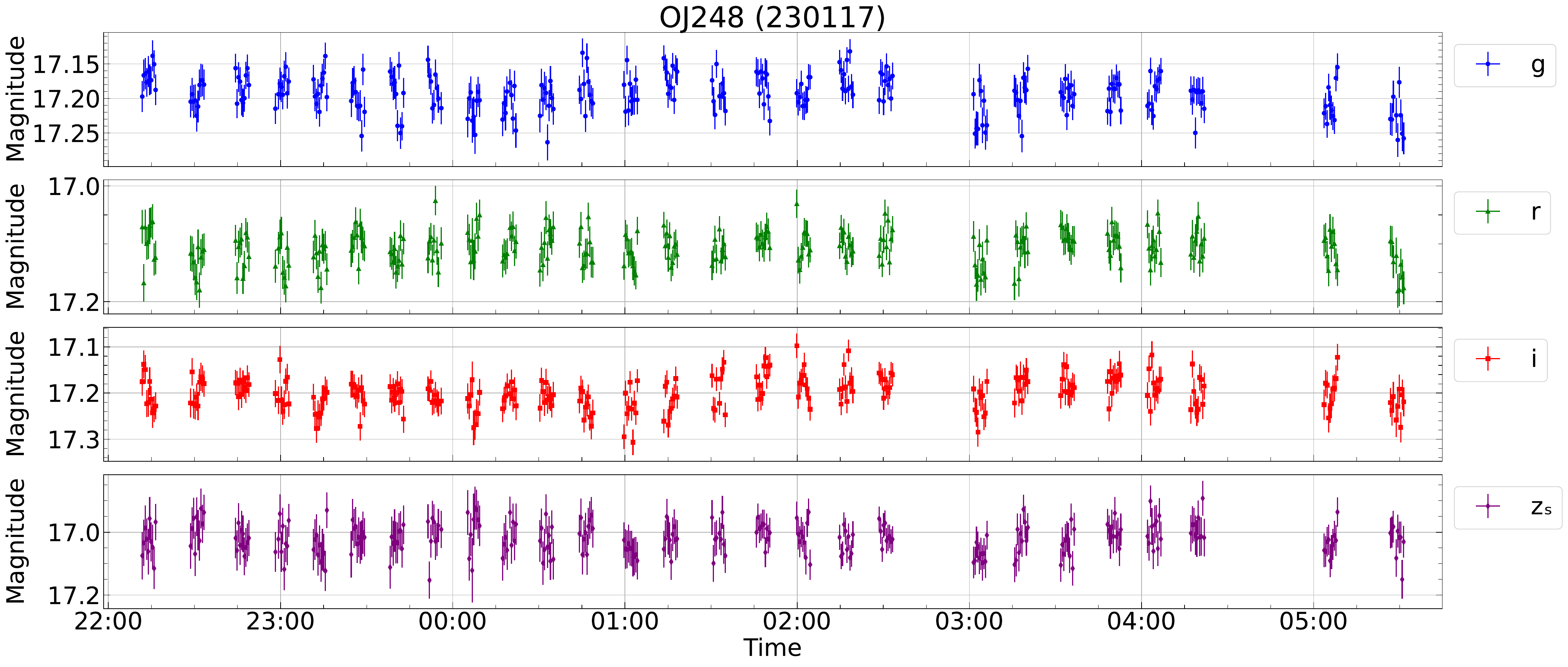}
\end{subfigure}
\hfill
\begin{subfigure}{\textwidth}
    \includegraphics[width=\textwidth,height=0.3\textheight]{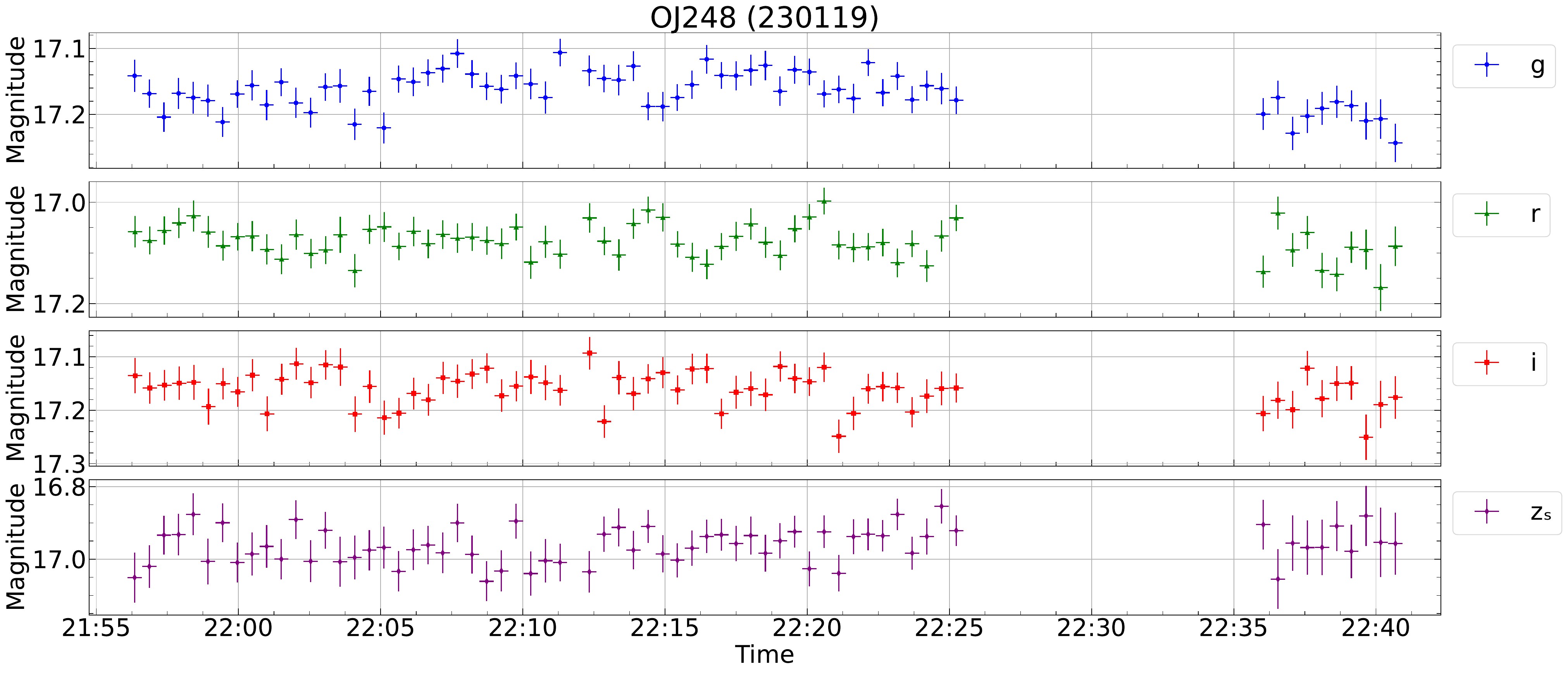}
\end{subfigure}
\caption{As Fig. \ref{txs0506+056_lc}, but for OJ248 on the nights of 2023 January 16, 2023 January 17, and 2023 January 19.}
\label{oj248_lc}
\end{figure*}

\begin{table*}
\centering
\begin{tabular}{cccccc}
\hline
Source       & Date   & $p$                 & $c$      & $F_\mathrm{enh}$ ($F_\mathrm{crit}$) & Variable? \\
\hline
TXS 0506+056 & 2023 Jan 15 & 4.1$\times10^{-3}$  & 0.25  & 1.11 (1.50)    & NV        \\
…            & 2023 Jan 16 & 0.34                & 0.08  & 1.41 (1.45)    & NV        \\
…            & 2023 Jan 18 & 0.21                & -0.13 & 1.43 (1.58)    & NV        \\
\hline
OJ287        & 2023 Jan 15 & 0.07                & -0.12 & 0.88 (1.53)    & NV        \\
…            & 2023 Jan 18 & 0.45                & -0.04 & 0.78 (1.31)    & NV        \\
…            & 2023 Jan 19 & 0.39                & -0.22 & 0.70 (2.79)    & NV        \\
\hline
PKS 0735+178  & 2023 Jan 15 & 4.6$\times10^{-3}$  & 0.33  & 4.66 (1.72)    & V         \\
…            & 2023 Jan 16 & 0.7                 & -0.08 & 1.84 (2.40)    & NV        \\
…            & 2023 Jan 17 & 3.9$\times10^{-10}$ & -0.35 & 3.13 (1.31)    & V         \\
\hline
OJ248        & 2023 Jan 16 & 0.22                & -0.24 & 0.41 (2.29)    & NV        \\
…            & 2023 Jan 17 & 0.25                & 0.07  & 1.15 (1.47)    & NV        \\
…            & 2023 Jan 19 & 0.35                & -0.12 & 0.69 (1.90)    & NV        \\
\hline
\end{tabular}
\caption{Colour variability statistics for each source on a given night. $p$ and $c$ refer to the Spearman rank correlation coefficients (significance and strength respectively), $F_\mathrm{enh}$ is the enhanced F-test value with the critical value as described in section \ref{sec: f-test}, and the final column describes whether or not the colour of the source was deemed variable on the given night. If $p<0.05$ and $F_\mathrm{enh}>F_\mathrm{crit}$ the source was deemed variable (V), otherwise not variable (NV).}
\label{colour_stats_full}
\end{table*}

\bsp	
\label{lastpage}
\end{document}